\documentclass[twocolumn]{aastex631}
\usepackage{amsmath}

\usepackage{graphicx} 
\usepackage{longtable}
\usepackage{natbib}
\usepackage{amssymb}
\usepackage{float}
\usepackage{makeidx}
\usepackage{morefloats}
\usepackage{captcont}
\usepackage{tabularx}

\begin{document}

\title{The Plane Quasar Survey: First Data Release}

\author[0000-0002-0355-0134]{Jessica Werk}
\affiliation{Department of Astronomy, University of Washington, Seattle, WA 98195, USA}
\affiliation{Astronomy and Astrophysics Department, University of California, San Diego, CA, 92093, USA}
\author[0000-0003-0789-9939]{Kirill Tchernyshyov}
\affiliation{Department of Astronomy, University of Washington, Seattle, WA 98195, USA}
\author[0000-0002-7483-8688] {Hannah Bish}
\affiliation{Space Telescope Science Institute, 3700 San Martin Drive, Baltimore, MD 21218, USA}
\author[0000-0003-4158-5116]{Yong Zheng}
\affiliation{Department of Physics, Applied Physics and Astronomy, Rensselaer Polytechnic Institute, Troy, NY 12180, USA}
\author[0000-0002-1129-1873]{Mary Putman}
\affiliation{Department of Astronomy, Columbia University, New York, NY 10027, USA}
\author[0000-0003-4797-7030] {Joshua Peek}
\affiliation{Space Telescope Science Institute, 3700 San Martin Drive, Baltimore, MD 21218, USA}
\author[0000-0003-2666-4430]{David Schiminovich}
\affiliation{Department of Astronomy, Columbia University, New York, NY 10027, USA}
\begin{abstract}

We present a sample of 305 QSO candidates having $|b| < 30^\circ$, the majority with GALEX magnitudes $NUV$ $<$ 18.75. To generate this sample, we apply UV-IR color selection criteria to photometric data from the Ultraviolet GAlactic Plane Survey (UVGAPS) as part of GALEX-CAUSE, the Million Quasars Catalog, Gaia DR2, and Pan-STARRS DR1. 165 of these 305 candidate UV-bright AGN (54\%) have published  spectroscopic redshifts from 45 different surveys, confirming them as AGN. We further obtained low-dispersion, optical, longslit spectra with the APO 3.5-m, MDM 2.4-m, and MDM 1.3-m telescopes for 84 of the candidates, and confirm 86\% (N = 72) as AGN, generally with $z < 0.6$. These sources fill a gap in the Galactic latitude coverage of the available samples of known UV-bright QSO background probes.   Along with a description of the confirmed QSO properties, we provide the fully-reduced, flux and wavelength-calibrated spectra of 84 low-latitude QSOs through the Mikulski Archive for Space Telescopes.  Future HST/COS spectroscopy of these low-Galactic-latitude QSOs has the potential to transform our view of the Milky Way and Local Group circumgalactic medium. 

\end{abstract}

\section{Introduction}
\label{sec:intro}

 
 With the recent public data releases from the {\emph{Gaia}} mission, we now have constraints on positions and velocities for nearly one billion of the Milky Way's stars. This spatial and velocity information from Gaia is especially powerful in combination with benchmark, ground-based surveys such as RAVE \citep{rave}, GALAH \citep{galah},  APOGEE-2 \citep{apogee2}, and soon, LSST \citep{lsst}.  Along with data from all-sky surveys covering different wavebands --  e.g. HI4PI \citep{HI4PI}, WHAM \citep{WHAM}, Planck \citep{planckdust}, and WISE \citep{WISE} --  it is now possible to map three-dimensional stellar age distributions, space motions, and metallicities along with three-dimensional distributions of dust, multiphase gas, and magnetic fields.  
 
  
Despite impressive progress in building the large datasets necessary for a comprehensive understanding of the Milky Way's evolution, knowledge of the content of our Milky Way's circumgalactic medium (MWCGM) beyond $\sim$15 kpc remains limited.   The MWCGM contains the byproducts of billions of years of star formation, drives its ongoing evolution, and harbors its fuel for future star formation. Unfortunately, it is also one of the most difficult Galactic components to study because we are embedded in a much denser ISM that effectively obscures our view, and because observations of our primary probes of its content, UV-bright background QSOs and halo stars, are biased toward high Galactic latitudes \citep[e.g.][]{zheng15, bish21}.   

The inner-CGM ($d\lesssim15$ kpc) of the Milky Way at Galactic latitudes  $|b| \gtrsim 30^{\circ}$ is well-studied. HI 21-cm emission maps of our own Galaxy's halo reveal giant arches, shells, and complexes of  neutral clouds moving with 35 $<  |$v$_{\rm LSR}| <  $~250 km/s, many with distance constraints that place them within the inner halo \citep{putman12}.  At distances $\lesssim$ 15 kpc, ionized diffuse high-velocity gas is distributed over large spatial scales, with a covering fraction of 67\%. This gas provides enough potential material to sustain the current level of star-formation in the Milky Way for at least another Gyr \citep{lehnerandhowk11, werk19}.  The velocities and surface densities of this material are generally thought to be consistent with gas cycling in a Galactic fountain, in which hot, over-pressurized gas flows from the disk into the halo, and subsequently cools, recombines and rains back down onto the disk \citep[e.g.][]{shapiro76, fraternalibinney}. Additionally, there is  strong evidence for a more localized, high-energy biconical wind emanating directly from the Galactic Center, observed in emission in hard and soft X-rays, in absorption at UV wavelengths,  in polarized radio emission, and in $\gamma$ rays, referred to as the Fermi Bubble \citep{snowden97, su2010, carretti13, bordoloi17, ashley22}.   

Beyond this very active disk-halo interface region, there is likely a CGM that extends beyond Galaxy's virial radius,  similar to other L* galaxies \citep[e.g.][]{prochaska11, zheng19, wilde21, wilde23}, which contains at least as much mass as in the stars and ISM in the disk itself \citep[e.g.][]{tumlinson11, werk14}. Apart from the massive Magellanic Stream \citep{2016ARA&A..54..363D}, we have yet to definitively detect this more extended component, although not for lack of trying. At Galactic latitudes $|b| > 25^{\circ}$, a QSO-star differencing technique reported significantly less CIV in the CGM of the MW compared to its low-redshift L* counterparts \citep{bish21}. This sensitive measurement consisted of subtracting unsaturated absorption lines of the CIV $\lambda\lambda$1548, 1549 doublet toward halo stars with known distances, well above the disk, from the same measurements toward distant QSOs nearby in projection to the stars ($<$ 2$^{\circ}$).  The spectral differencing technique, despite being sensitive to low column densities at v $\sim0$, finds mostly {\emph {upper limits}} on the ionized content of the $>$ 10 kpc MWCGM. However, because all of the Quastar Survey's lines of sight lie at $|b| > 25^{\circ}$,  one option remains to resolve this potential Milky Way anomaly: a very extended, ionized gas disk as has been proposed by \cite{bregman18}, only visible at low Galactic latitudes. 

A major limitation in studies of the Milky Way CGM is a dramatic shortage of UV-bright QSO sightlines at Galactic latitudes $|b| < 25^{\circ}$, {\emph{representing a huge gap that covers approximately half of the sky.}  Of the more than 1000 known UV-bright QSOs, only about $\sim$5\% of them have Galactic latitudes $|b|$ $<$ 25$^\circ$, and there is only one previously-detected UV-bright QSO confirmed at $|b|$ $<$ 10$^\circ$ \citep{uvqs}.  Similarly, of the hundreds of QSO lines-of-sight with high-resolution, high-quality {\emph{HST}}/COS spectra available in the archives, there are only 8 with $|b| < 25^{\circ}$. UV-bright QSOs near the plane of the Milky Way are notoriously difficult to find because of obscuration by dust and neutral gas, and the lack of deep UV surveys in this area until very recently. 

Low-$|b|$ QSOs are critical probes of the structure and kinematics of any extended disk feature that may contribute to the content of the MWCGM.  Our Plane Quasar Survey (PQS) addresses this major limitation in our absorption-line studies of the Milky Way by providing a sample of 305 low-Galactic-latitude, UV-bright QSO candidates. In this paper, we confirm $>$ 60\% of these candidates as AGN and provide redshift estimates. Section \ref{sec:select} presents our two selection techniques to identify these low Galactic latitude, UV-bright QSO candidates; Section \ref{sec:obs} details the spectroscopic observations for redshift confirmation; and Section \ref{sec:redshifts} provides details on our two methods of deriving redshifts for confirmed QSOs. All of our spectra and tables are available via MAST as a High Level Science Products via \url{https://doi.org/10.17909/86my-mr86}\footnote{\url{https://archive.stsci.edu/hlsp/pqs/}}.


\section{Candidate QSO photometric selection}\label{sec:select}

We use two primary selection techniques to identify low-Galactic-latitude UV-bright QSO candidates. The first uses $NUV$ data from the Ultraviolet GAlactic Plane Survey (UVGAPS; Mohammed et al. 2019\nocite{mohammed19}), which produced a high resolution map of the Milky Way's Galactic plane in the $NUV$ using the Galaxy Evolution Explorer \citep{GALEX}. UVGAPS covers an area of $\sim7200$ square degrees (360 degrees $\times$ 20 degrees) with a full width half max resolution of 4.5-6$\arcsec$, with 2$\arcsec$ pixels, which is a larger footprint at a higher resolution than previous UV all-sky surveys within the same region. We cross-matched the $NUV$ source catalog of UVGAPS with the candidate QSOs presented in the Million Quasars Catalog (MilliQUAS; Flesch et al. 2023\nocite{milliquas}) to identify candidate UV-bright QSOs, all having $|b| < 10^\circ$ and $NUV < 20.0$. The objects in the MilliQUAS catalog have been selected as likely QSOs using mid-IR color-color classifications from WISE all-sky survey data. Finally, we use {\emph{Gaia}} parallax data for all targets to exclude those that are likely Galactic stars by imposing a parallax cut of $<$ 0.5 milliarcseconds. In total, this selection method yields 83 Galactic plane quasar candidates. 

We supplemented this very low Galactic latitude sample by imposing WISE photometric selection criteria on a broader range of targets from GALEX, similar to the approach of the UVQS Survey \citep{uvqs}.   Our candidate sample selection differs from that of UVQS in that we do not use the $FUV$ criteria.  Doing so would bias us against targets in the plane of the Milky Way (which UVQS does by design) as the plane was primarily surveyed with GALEX CAUSE when GALEX $NUV$ was the only detector available. For this supplementary sample, we select targets from the AllWISE catalog \citep{allwise} to have $SN > 5$ in $W1$ and $W1 - W2 > 0.6$, reject nearby stars by imposing a cut on  {\emph{Gaia}} parallax that requires candidates to have $p < 0.5$ milliarcseconds, and finally select targets with GALEX $NUV < 18.75$ (see \ref{fig:colormag}). We then examined the distribution of $NUV - W2$ colors of the overlapping UVQS-spectroscopically-confirmed QSOs, and imposed an additional color cut of $NUV - W2 > 5$ after discerning that 96\% of that sample had $NUV - W2$ colors redder than that limit. Finally, we very conservatively require that our candidates do not exhibit $>$ 6 magnitudes of extinction in the $NUV$ \citep{sfd}. Our final selection criteria for this secondary Plane QSO sample is that $|b| < 30^\circ$. This selection method yielded 222 total QSO candidates. In addition, we note that this supplementary candidate selection does identify 73 (of the 222) QSO candidates that have been previously confirmed as AGN by UVQS.

 The coordinates and photometric properties of our complete sample of 305 Plane QSO candidates are provided in Table 1. 165 of these PQS candidates are spectroscopically confirmed as AGN from 45 different spectroscopic surveys, and the quoted spectroscopic redshifts and references, when available, are also provided in Table 1. Because the data come from a variety of surveys,  with varying spectral properties and redshift accuracy, we refer readers to the original source papers for details.  Generally, the lowest Galactic latitude PQS candidates from the UVGAPS selection method are the ones without archival redshifts, and we prioritized these for our own spectroscopic followup, described in subsequent sections. However (somewhat accidentally), we re-observed 45 of these sources with existing spectroscopic data. In part, this was due to not completing a full archival cross-matching of our candidates before commencing our observing runs. These re-observed targets allow us to compare QSO redshifts derived from different telescopes and instruments, which we discuss in Section \ref{sec:redshifts}. 
 \nocite{1953GCRV..C......0W, 1989ApJS...69....1H,
1990AJ....100.1274W,
1991A&A...246L..24M,
1991ApJS...76..813S,
1993AJ....105..660W,
1995A&A...300..323J,
1995A&AS..110..469B,
1999A&AS..134..483H,
1999A&AS..139..575W,
2000ApJ...543..552S,
2000ApJS..129..547B,
2001AJ....121.2843B,
2001ApJ...546..744N,
2002LEDA.........0P,
2004ApJ...617..192M,
2004MNRAS.355..273C,
2006A&A...459...21M,
2008A&A...482..113M,
2009ApJS..182..543A,
2009MNRAS.399..683J,
2011MNRAS.414..500H,
2011MNRAS.416.2840L,
2011NewA...16..503M,
2012A&A...545A..15S,
2012AJ....143...64J,
2012ApJ...751...52E,
2012ApJS..203...21A,
2013ApJ...767...14P,
2013MNRAS.430.2464M,
2014A&A...561A..67P,
2015A&A...573A..59G,
2015ApJS..219...12A,
2015RAA....15.1438H,
2016ApJ...818..113N,
2017AJ....154..114O,
2017ApJS..233....3T,
2018A&A...613A..51P,
2018AJ....155..189D,
2018ApJ...861...49H,
2018yCat.1345....0G,
2020AJ....160..120J,
2020ApJS..250....8L}
  
  In Figure \ref{fig:colormag}, we show a color-magnitude plot of $WISE$ and $GALEX$ photometry for the 1055 confirmed AGN in UVQS and for the 305 Plane QSO candidates whose selection is described above. Our set of UVQS-like photometric selection criteria are shown as dashed lines. The $WISE$ magnitudes are drawn from the allWISE Source catalog (``w2mpro" and ``w1mpro"), and are measured from profile-fitting photometry, while the $GALEX$ $NUV$ magnitudes are drawn from the band-merged source catalog GR6 data on MAST. The distribution of both samples within this $WISE - GALEX$ color magnitude diagram overlaps considerably. Every candidate and confirmed source that lies outside of these bounds originates in the first candidate selection method that uses UVGAPS data and the MilliQUAS catalog. Overall this sample does not appear markedly different from the sample selected from the second-set of criteria described above. The dark purple diamonds in Figure \ref{fig:colormag} show our Galactic plane QSO candidates that are spectroscopically confirmed as AGN, while the dark maroon stars show candidates that exhibit stellar spectra. Light purple x's show Plane QSO candidates that remain unobserved.  
  
\begin{figure}[t!]
\begin{center}
\includegraphics[width = 3.4in]{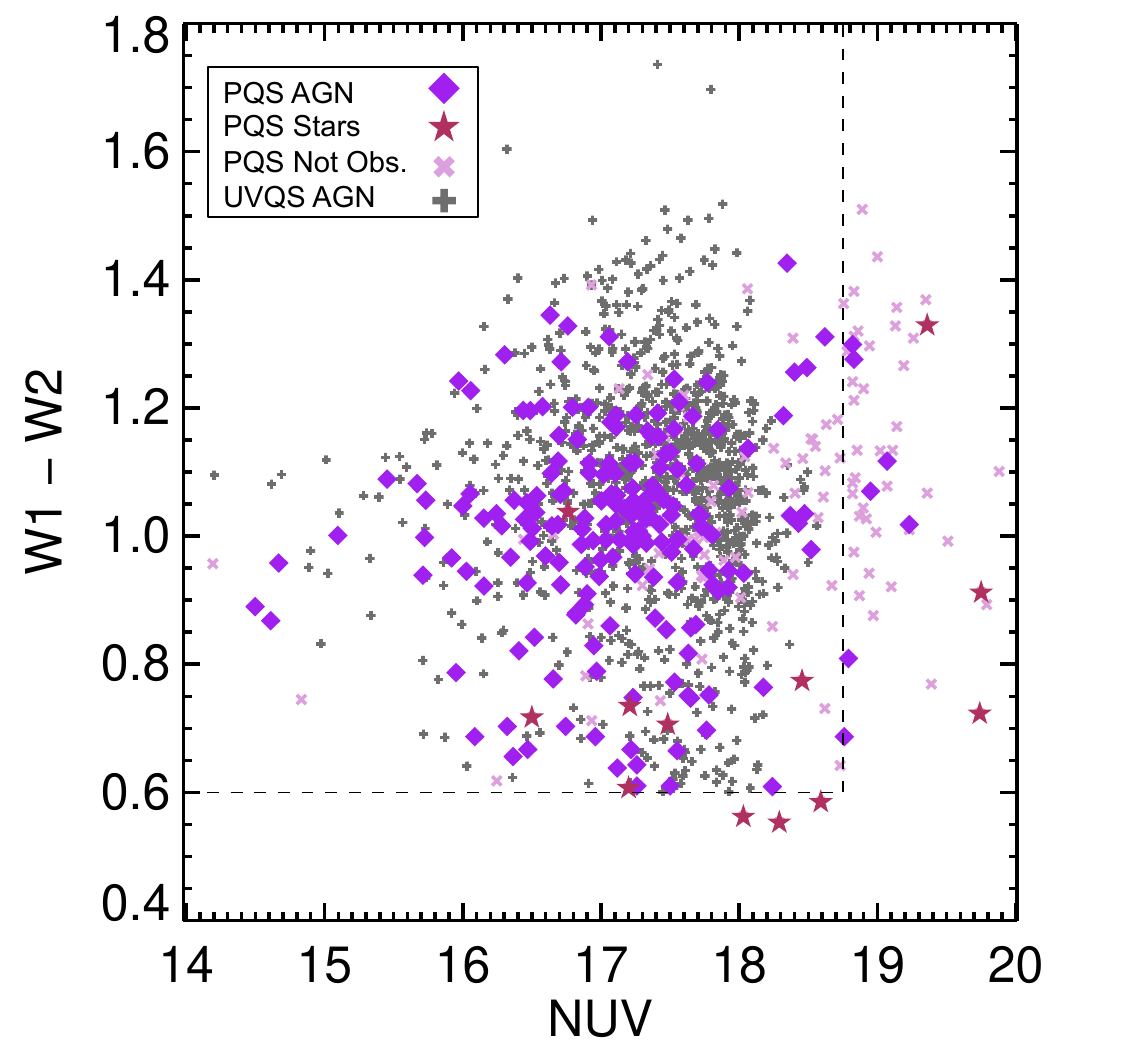}
\end{center}
\caption{ {{{\sc UV-bright QSO Photometric Selection:} A WISE-GALEX color-magnitude diagram showing the distribution of different UV-bright QSO samples in GALEX NUV magnitude vs $W1 - W2$ color,  including the our Plane QSO candidates from both selection methods (dark and light purple shapes). The gray plus signs show the confirmed AGN from the UVQS Survey \citep{uvqs}, including those that were previously identified by the Milliquas Survey \citep{milliquas}. Purple diamonds highlight the spectroscopically-confirmed low-latitude QSOs, which fill a gap in the existing datasets. The 12 maroon stars show objects in our candidate list that are spectroscopically confirmed  as stellar sources, and the light purple x's are targets from our candidate list that have not yet been observed.  }} }
\label{fig:colormag}
\end{figure}

Figure \ref{fig:molleweide} shows the all-sky distribution in Galactic coordinates of our sample of candidates, as well as the UVQS confirmed AGN and AGN with high-SN UV spectroscopy in the {\it{Hubble Spectroscopic Legacy Archive}} \citep{hsla}. Compared to previously-catalogued UV-bright AGN, our Plane QSO candidates, by design, generally lie at lower Galactic latitudes and are most numerous at Galactic longitudes $l > 90^\circ$. The dark purple diamonds and stars in Figure \ref{fig:molleweide} show the 84 objects for which we have obtained spectra and those objects with archival spectroscopic redshifts available. Our initial data release for the Plane QSO survey represents a spectroscopic completeness of $\sim$63\%. Given the finite lifetime of $HST/COS$, we believe the public release of this limited dataset to be both important and timely, especially given that many of the targets lie in areas of interest completely devoid of existing high-resolution UV spectroscopy. 

\begin{figure*}[t!]
\begin{center}
\includegraphics[width = 7.4in]{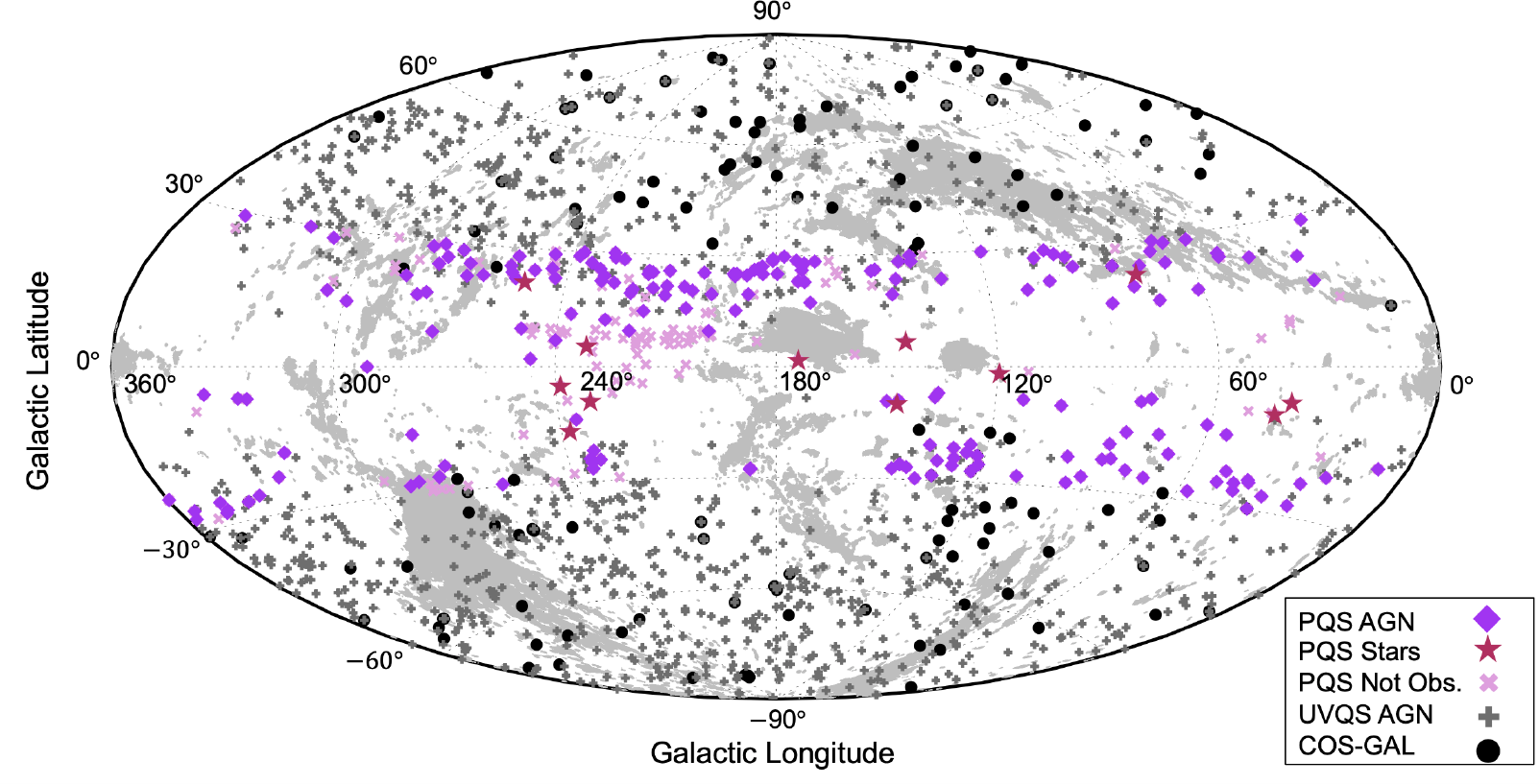}
\end{center}
\caption{ {{{\sc UV-bright QSO All-Sky Distribution:}  An all-sky Aitoff projection in Galactic coordinates, centered at $b = 0^{\circ}$, $l = 180^{\circ}$, showing the distribution of different UV-bright QSO samples, including the Plane QSO candidates (dark and light purple shapes).  For orientation within this frame of reference, the underlying gray map shows high-velocity ($|v_{\rm LSR}| > 80$ km s$^{-1}$) HI detected in 21-cm emission from the HI4PI Survey \citep{HI4PI, 2018MNRAS.474..289W}, generally with N$_{\rm HI} >$ 10$^{18}$ cm$^{-2}$.  The black circles show the 401 archival QSO sightlines with high-resolution {\emph{HST}}/COS spectra from the COS-GAL database \citep{zheng19} , and the gray plusses show all the confirmed AGN from the UVQS Survey \citep{uvqs}, including those that were previously identified by the Milliquas Survey \citep{milliquas}. Purple diamonds highlight the spectroscopically-confirmed low-latitude QSOs, which fill a gap in the existing datasets. The 12 stars show objects in our candidate list that were confirmed stellar sources, and the light purple x's are targets from our candidate list that have not yet been observed.  }} }
\label{fig:molleweide}
\end{figure*}

\section{Observations and Data Processing}
\label{sec:obs}
In total, we obtained spectra of 84 Plane QSO candidates during the 2020 calendar year.  For the majority of the candidates we observed, we obtained low-dispersion, longslit optical spectra using the APO 3.5-m telescope with the $DIS$ spectrograph and with the MDM 2.4-m telescope with the OSMOS spectrograph.  For six candidates, we obtained spectra on the MDM 1.3-m telescope using CCDS. We generally prioritized targets with the lowest latitudes and brightest UV magnitudes such that they would be ideal sources for {\emph{HST}}/COS follow-up to kinematically isolate any extended, ionized features in the Galactic plane. In addition, we prioritized potential lines of sight at Galactic longitudes that are optimal for isolating extended structures in a rotating disk, observed 8 kpc from its center. Specifically, we looked primarily at Galactic longitudes that are looking away from Galactic center ($l = 0^{\circ}$) and anti-center ($l=180^{\circ}$), and either toward or against the direction of rotation, respectively $l \approx 90^{\circ}$, and  $l \approx 270^{\circ}$. This section details the observations on each telescope, and describes the properties of the reduced, one-dimensional, calibrated spectra. 

All of our calibrated, 1D spectra are available at MAST as a High Level Science Product via \url{https://doi.org/10.17909/86my-mr86 }\footnote{\url{https://archive.stsci.edu/hlsp/pqs/}}.  Given the range of instruments employed, and the diversity of targets, the resultant spectra are of varying quality. The S/N ranges from 3 $-$ 100. We note that every target we observed has a spectrum of sufficient quality to determine a source type (e.g. star or AGN), as well as a redshift if the target was an AGN.

\subsection{Optical Spectroscopy with the APO 3.5-m}
\subsubsection{Description of Observations}
APO data were taken under largely clear skies with variable atmospheric conditions over the course of four grey-time half-nights in February, October, and December of 2020. We used the {\emph{Dual Imaging Spectrograph}} (DIS), which has a spatial scale of $\sim$0.4$\arcsec$/pixel, with the R300 grating in the red channel. All spectra were taken with a 1.2\arcsec~ slit, which was the approximate seeing on each night we were able to observe.  Our APO red-channel instrumental setup results in a linear dispersion of 2.31\AA/pixel and a spectral coverage of 4620\AA~about a central wavelength of 7500\AA. During our runs,  the blue channel of DIS was not functional so we were only able to use the red channel. Each candidate was observed with a 600-second exposure time.

\subsubsection{Data Reduction and Final Products}

The APO two-dimensional spectra were reduced and calibrated using a custom python package {\emph{pydis}}\footnote{\url{https://github.com/StellarCartography/pydis}} developed by J. Davenport. The spectra are bias-subtracted, flat-fielded using quartz lamp images in the same spectral configuration as the data, and wavelength-calibrated using Hg-Ne-Ar arc-lamp exposures. The wavelengths were then shifted to a vacuum and heliocentric reference frame. The wavelength solutions have a typical rms of 0.8 \AA. The targets were extracted to one-dimensional spectra using an optimized aperture, and were flux-calibrated in a relative sense using observations of a spectrophotometric standard star taken during twilight each night. The flux calibration is coarse -- we have not corrected for slit-losses and only observed the standard star during twilight on nights during which the conditions were highly variable. Nonetheless, the calibration that is applied should allow for an approximate comparison of emission-line fluxes in QSO spectra in a relative sense. The average S/N of each spectrum is 18 per resolution element, and each spectrum is high-enough quality to allow us to determine the source-type, and redshift if it is extragalactic.

\subsection{Optical Spectroscopy with the MDM 2.4-m and 1.3-m}
\subsubsection{Description of Observations}
MDM 2.4-m spectra were obtained with the Ohio State Multi-Object Spectrograph (OSMOS) in queue-mode observing from November - December 2020. OSMOS has a spatial scale of $\sim$0.3$\arcsec$/pixel.  Every target was observed for 600s with the MDM4K detector, a 1.2\arcsec~slit,  and the VPH blue grism with $R=1600$ (dispersion $\approx$ 0.7\AA/pixel). The grism uses an inner and center slit with wavelength coverages of 3900 $-$ 6800 \AA~(inner) and 3100 $-$ 5950 \AA (center). Three of our targets were observed remotely on the 1.3-m with the Boller and Chivens CCD Spectrograph (CCDS) in February and March of 2020 under moderately unfavorable atmospheric conditions. On the 1.3-m, CCDS has a spatial scale of $\sim$0.4$\arcsec$/pixel. For these targets, we exposed for 600s using a 1.5\arcsec~slit, and used the 150 l/mm grating with a central wavelength value of 5500 \AA, which has a dispersion of $\sim$3.3\AA/pixel.

\subsubsection{Data Reduction and Final Products}

MDM spectra were reduced using the IRAF  {\it{specred}} package \citep{IRAF, specred}, which includes bias subtraction, flat-field calibrations using quartz lamp images, and wavelength calibration using Hg-Ne-Ar arc-lamp exposures.  We shifted all wavelengths to a vacuum and heliocentric reference frame. The wavelength solutions for the OSMOS data  have a typical rms of 0.4 \AA, whereas those of CCDS are 1.4 \AA . We performed the extraction of the 1D spectrum using the {\it{APALL}} task, and manually adjusted each aperture size in an interactive procedure to contain the majority of the target flux. We flux-calibrated the data using observations of several spectrophotometric standard stars, all observed during twilight time, generally to a relative accuracy of  $\sim$10\%.

\section{Redshift Determination}
\label{sec:redshifts}

\begin{figure*}[t!]
\begin{center}
\includegraphics[width = 6.8in]{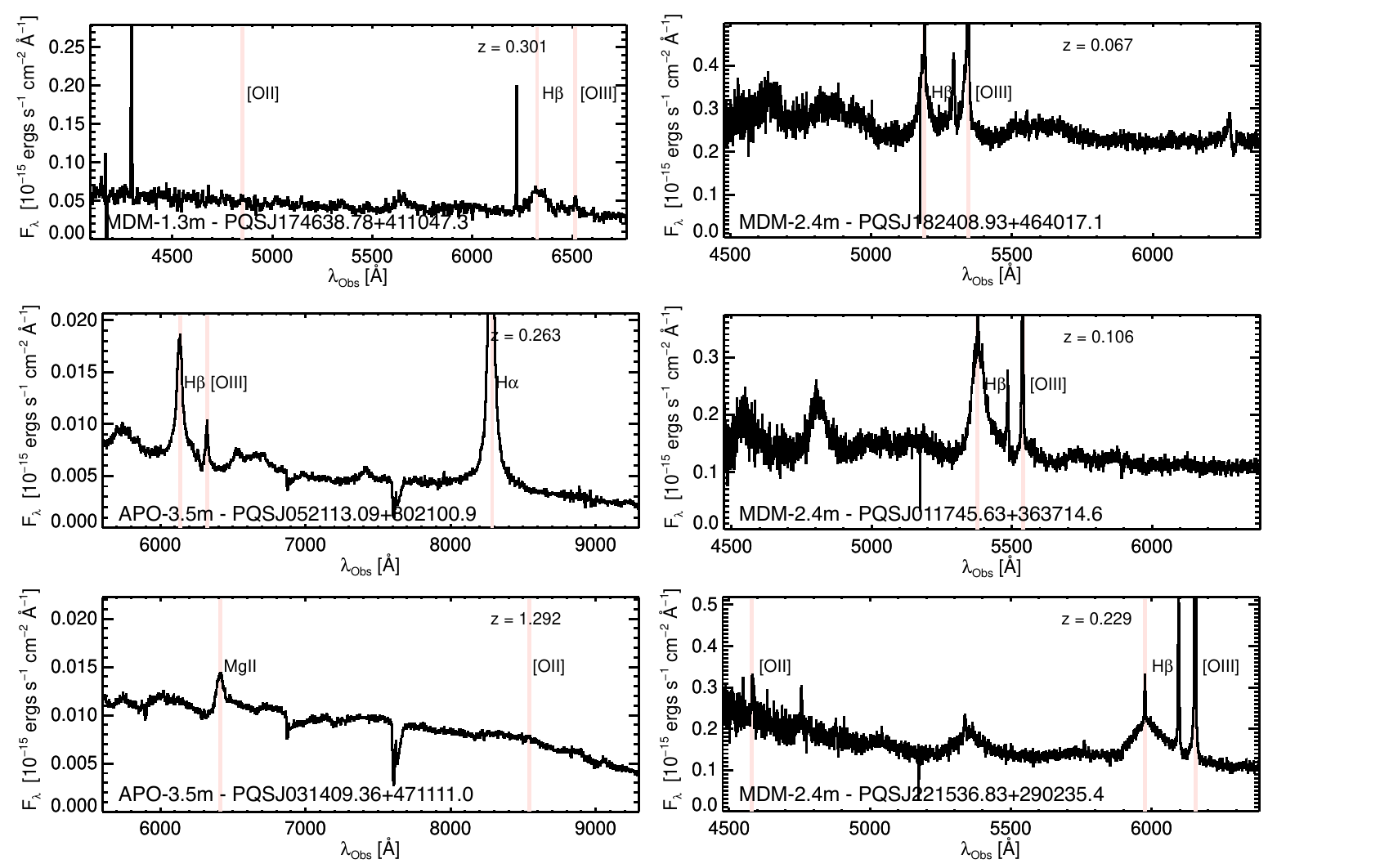}
\end{center}
\vspace{-0.2in}
\caption{ {{{\sc Example PQS Spectra:} We show six flux and wavelength-calibrated, 1D spectra from the three different telescopes used during our PQS observing runs. On the upper right of each panel, we provide the MARZ redshift determined using the template QSO spectrum fits. Some of the strong emission lines identified in each spectrum are labeled, and the telescope and PQS target name are given in the lower right hand corner of the spectra. All of the spectra are available to download on MAST. }} }
\label{fig:spectra}
\end{figure*}

The primary goal of this paper is to provide spectroscopic confirmation of low-$b$, UV-bright QSO candidates and to determine an approximate redshift. Visual examination of our 84 reduced QSO candidate spectra revealed that 12 of these candidates are very likely stars, with spectra generally consistent with G and M dwarfs. We list these stars and the telescopes at which their spectra were taken, in Table 2, along with the 72 confirmed AGN sources. The majority of the confirmed stellar sources (8/12) come from our UVGAPS selection, and tend to lie at $|b| \lesssim 10^{\circ}$ (11/12). We note that our UVGAPS selection technique yielded a lower AGN-confirmation rate (9/17; 53\%) than our UVQS-like sample selection method (63/67; 94\%), likely because the former sample is more concentrated along the Galactic plane, where stellar contamination will naturally be higher. We consider the stars no further in our analysis. 

 Uncertainties in deriving systemic redshifts from broad emission lines in low-dispersion optical spectra of QSOs often exceed 500 km s$^{-1}$, and individual redshift measurements for the same QSO can differ by up to 1000 km s$^{-1}$, depending on which lines are used in the fitting algorithm \citep[e.g.][]{SDSSQSOs}. By comparison, errors from wavelength calibration alone are $\sim$ 50 km s$^{-1}$ for our low resolution optical spectra. Well-tuned methods of template spectral fitting and spectral masking can reduce QSO redshift errors by a factor of twenty \citep{hewett2010}, but again, depend on the specific emission lines that are present in the spectrum. For example, using narrow [OII] and other low-ionization lines for redshift measurements may offer significant advantages over [OIII]-based redshift measurements, which often show signs of outflows and shifts of up to 400 km s$^{-1}$ with respect to stellar absorption lines \citep[e.g.][]{boroson05, shen16}. With these caveats and realities in mind, we employ two methods to estimate the redshifts of our confirmed AGN sources with heterogenous spectral quality and wavelength coverage. When possible, we compare our independently-derived redshift measurements with those available in the literature for the same QSOs. 

 Figure \ref{fig:spectra} shows six one-dimensional, flux and wavelength calibrated PQS spectra of varying quality, with at least one from each of the telescopes we used for spectroscopic confirmation. Identified emission lines are labeled in each panel, and the redshift of each source is labeled in the upper right corner of each spectrum. Note, the spectra have not been cleaned of cosmic rays and are not corrected for telluric absorption. Because the targets span a range of redshifts, and were observed using different instruments and set-ups,  we see a variety of emission lines in each spectrum. Generally, all of our spectra contained at least one feature that allowed us to determine its spectral type and redshift.

\begin{figure*}[ht!]
    
\begin{center}
\includegraphics[width = 5.1in]{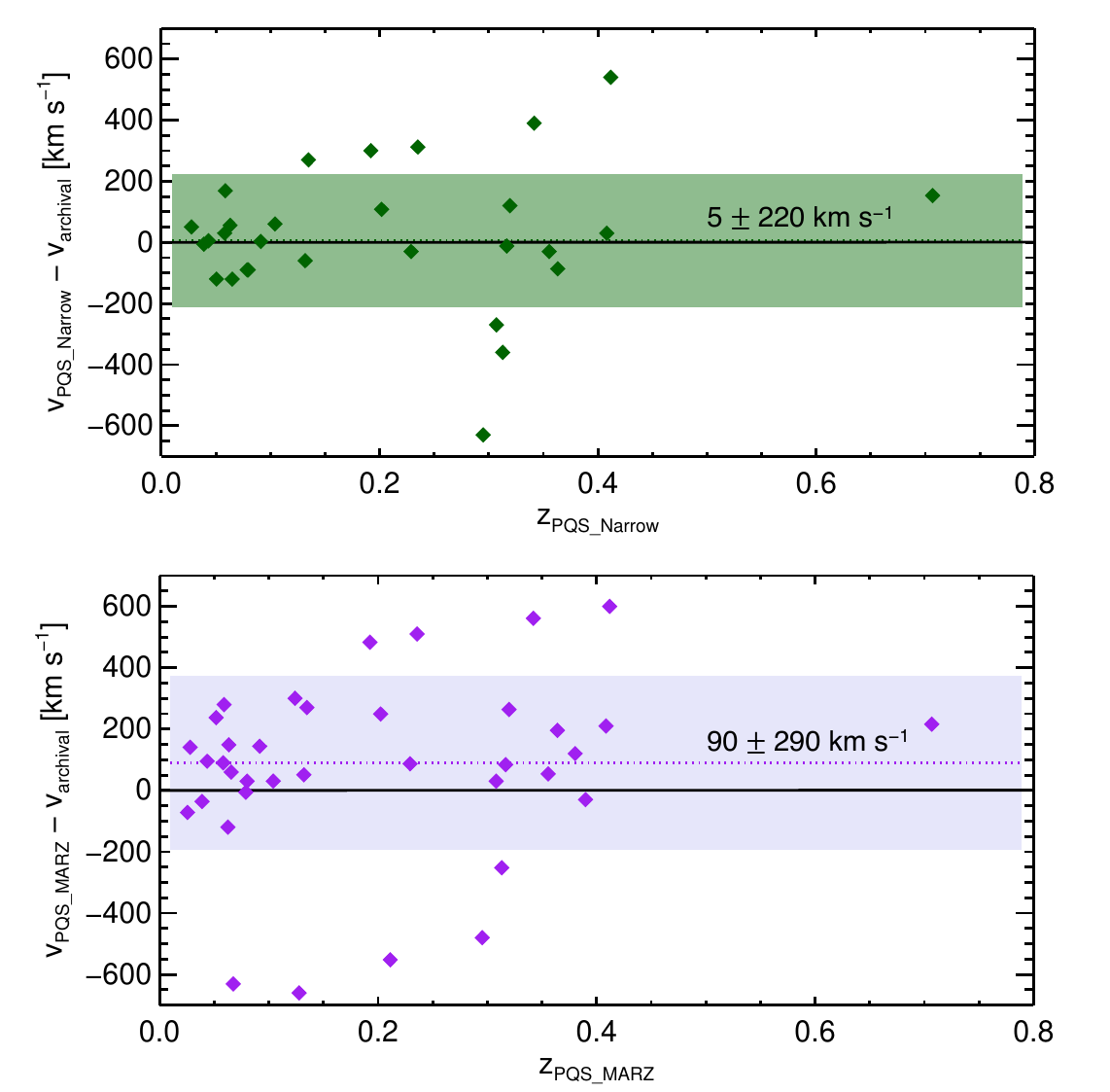}
\end{center}
\caption{ {{{\sc Top:} We compare our redshifts derived from narrow-line fits (generally to [OIII]  $\lambda \lambda$ 4960, 5008) with those available in the literature, from different sources, determined with a variety of methods based on data of highly variable quality.   {\sc Bottom:}  We compare our MARZ template-fitted QSO redshifts with those same literature QSO redshift estimates. On both panels, we have labeled the mean velocity offsets between the PQS redshifts and archival estimates with $\pm$ 1$\sigma$ spreads. We have eliminated from this analysis the eight cases discussed in the text with $>$ 1000 km s$^{-1}$ differences.   }} }
\label{fig:compz}
\end{figure*}

In our first redshift estimation, we used MARZ, Manual and Automated Redshifting Software \citep{Hinton16}. We interactively fit a template QSO spectrum to our 1-D, flux and wavelength-calibrated spectra, and in approximately 30\% of the cases, the automatic algorithm finds a high-quality redshift that appears to match the QSO features in the spectra (rather than, e.g.,  a spurious cosmic ray or a poorly-subtracted sky line). We hypothesize that the automated fitting routine fails in the majority of cases because our PQS spectra are uncleaned for cosmic rays and telluric absorption;  both appear to be affecting the MARZ automated algorithm.  In the majority of our spectra, we have to manually adjust the redshift to our ``by-eye" estimate based mostly on line identifications of  the Balmer series lines and [OIII] $\lambda$ 5008, and then re-run the template fitter. The variance in redshift fitting results reported by \cite{Hinton16} is generally very low ( $\Delta z / (1+ z)$ $\approx$ 3 $\times 10^{-4}$). Observational errors (e.g. wavelength calibration) and systematic errors (e.g. ubiquitous QSO outflows) will dwarf these fitting errors. As discussed above, the errors on our redshifts are likely well above 100 km s$^{-1}$. 

Our second redshift estimation is made using Gaussian fits to only the narrow emission lines present in the QSO spectra. For approximately 25\% of the sample, only broad Balmer lines or the MgII doublet are detected, and so we cannot employ this secondary redshift estimate. In the majority of cases in which narrow lines are present, we use the [OIII] lines at $\lambda \lambda$ 4960, 5008. The [OII] doublet at $\lambda \lambda$ 3727, 3729 is blended together at these low spectral resolutions, and is often in a very noisy region of the spectrum and thus not often well-detected. \cite{hewett2010} find systematic offsets of [OIII]-derived redshifts of 45 $\pm$ 5 km s$^{-1}$ to the blue in comparison to centroids from the CaII K absorption line at 3934.8 \AA, however we do not apply this shift to our narrow-line redshift estimates.

We provide both of these redshift estimates, in addition to the archival redshift estimate and its source, in Table 2. We compare our two redshift estimates with those values we find in our archival reference sample in the upper and lower panels of Figure \ref{fig:compz}. Given the variety of origins of the spectra, we are not implying that the reference redshifts are more or less correct than our values, but instead showing the typical variance in redshift estimates for the same sources. Our narrow-line redshifts agree reasonably with the previously published values, with a mean offset of only 5$\pm$ 220 km s$^{-1}$. Our MARZ-template-fitting-derived redshifts are biased high by 90 $\pm$ 290 km s$^{-1}$, but this is not significant considering we expect our QSO redshift errors to be hundreds of km s$^{-1}$.  

Not included in Figure \ref{fig:compz} are eight sources with redshift estimates that differ from the published values by more than 1000 km s$^{-1}$, well outside of our expected errors. We consider these sources highly discrepant. That said, the source of the large errors in the majority of cases is almost certainly rounding error. In 7/8 discrepant cases, the published redshifts are given to only two significant figures, which means differences larger than 1000 km s$^{-1}$ are expected. The remaining true discrepancy is PQSJ211452.57+060742.9, or 2MASS J21145258+0607423. The published redshift value for this QSO is 0.457 \citep{2017AJ....154..114O}, compared to our MARZ redshift of 0.4608, derived from two very broad emission lines we identified as MgII and H$\gamma$. \cite{2017AJ....154..114O} state the wavelength covered in the Keck  spectrum to be 3805 $-$ 5976 \AA, and thus, they would have detected only the broad MgII line for a redshift estimate. A redshift uncertainty of $\pm$ 1000 km s$^{-1}$ is therefore appropriate for this target. 


\section{Summary} 

In this paper, we have presented 305 low-Galactic-latitude, UV-bright QSO candidates that move us closer to unbiased coverage in Galactic coordinates of UV-bright background probes of the MWCGM. Of the 84 candidates we observed with low-resolution, ground-based spectroscopy from MDM and APO, we confirmed 72 as AGN. Combined with archival redshift estimates, our redshift estimates for these 72/84 (86\%) confirmed AGN bring us to 193 / 305 (63\%) spectroscopically-confirmed, UV-bright QSOs at low Galactic latitudes. When possible, we compare our redshifts with those available in the archives via a SIMBAD crossmatch and generally find good agreement to within $\pm$ 200$-$300 km~s$^{-1}$.  We make our full catalog, redshift estimates, and spectra publicly available via MAST as a High Level Science Product via \url{https://doi.org/10.17909/86my-mr86 }\footnote{\url{https://archive.stsci.edu/hlsp/pqs/}} in hopes that it will be used in future studies of the Milky Way and Local Group CGM while HST/COS is still available. 

\section{Acknowledgements}
   
  We thank the dedicated telescope support staff at both APO and MDM observatories, especially for making these remote observations possible during a global pandemic.  In addition, J.K.W. would like to thank the newly formed Department of Astronomy and Astrophysics at UC San Diego, especially Alison Coil and Dusan Keres,  for welcoming her to their beautiful campus as a Margaret Burbidge Visiting Professor during Winter 2024. J.K.W. is also deeply thankful for Laura Chomiuk and Kate Rubin, whose regular support through mutual accountability makes it possible to complete projects like this one while also juggling the responsibilities of being faculty.  This research has made use of the SIMBAD database, operated at CDS, Strasbourg, France \citep{2000A&AS..143....9W}, and the HSLA database, developed and maintained at STScI, Baltimore, USA. Support for this work was provided by NSF-CAREER 2044303. 
  
\facilities{McGraw-Hill, Hiltner, ARC}

\startlongtable
\begin{deluxetable*}{llcccccc}
\tablewidth{0pc}
\tablecaption{Plane QSO Candidates\label{tab:pqscandprops}}
\tabletypesize{\scriptsize}
\tablehead{\colhead{PQS Name}  & \colhead{SIMBAD Name} & \colhead{$l$}  &\colhead{$b$} &\colhead{m$_{\rm NUV}$} &\colhead{W1$-$W2} &\colhead{Ref z} &\colhead{Ref} } 
\startdata
PQSJ002324.76+605103.6 & IRAS 00206+6034 &  119.5240 &   -1.8350 & 16.50 &  0.72 &  &  \\
PQSJ004342.52+372520.1 & LEDA 2100384 &  121.2328 &  -25.4239 & 16.74 &  0.70 & 0.0800 & 2015RAA....15.1438H \\
PQSJ004818.93+394112.1 & 2MASS J00481898+3941116 &  122.2785 &  -23.1810 & 16.71 &  0.92 & 0.1340 & 2016ApJ...818..113N \\
PQSJ005050.78+353644.0 & 2MASS J00505073+3536429 &  122.7967 &  -27.2593 & 16.47 &  0.93 & 0.0580 & 2016AJ....152...25M \\
PQSJ010131.18+422935.1 & 2MASS J01013117+4229356 &  124.9137 &  -20.3397 & 16.94 &  0.99 & 0.1900 & 2016ApJ...818..113N \\
PQSJ010415.84+402243.5 & UVQS J010415.77+402243.9 &  125.5740 &  -22.4271 & 16.50 &  1.01 & 0.1910 & 2016AJ....152...25M \\
PQSJ011745.63+363714.6 & 2MASS J01174562+3637144 &  128.8032 &  -25.9506 & 17.51 &  1.03 & 0.1100 & 2015RAA....15.1438H \\
PQSJ011849.46+383618.8 & [VV2000] J011849.5+383619 &  128.7829 &  -23.9551 & 17.92 &  0.92 & 0.2200 & 2015RAA....15.1438H \\
PQSJ012400.58+410709.9 &  &  129.5100 &  -21.3358 & 17.57 &  1.21 &  &  \\
PQSJ013741.29+330934.9 & 3C  48 &  133.9625 &  -28.7194 & 17.33 &  1.06 & 0.3690 & 2016AJ....152...25M \\
PQSJ013925.09+370454.9 & LEDA 2092626 &  133.4563 &  -24.8072 & 17.39 &  0.87 &  &  \\
PQSJ015536.05+311518.3 & LEDA 3095555 &  138.7027 &  -29.6440 & 17.63 &  0.82 & 0.1334 & 2020ApJS..250....8L \\
PQSJ015756.87+402413.3 & UVQS J015756.83+402413.1 &  136.4040 &  -20.7216 & 16.70 &  1.16 & 0.4580 & 2016AJ....152...25M \\
PQSJ021248.58+330643.3 & ATO J033.2024+33.1119 &  141.9604 &  -26.7352 & 17.69 &  1.11 &  &  \\
PQSJ021935.13+534019.9 &  &  135.8717 &   -6.9702 & 17.26 &  0.61 &  &  \\
PQSJ022206.31+522106.1 & 2MASS J02220630+5221059 &  136.6858 &   -8.0820 & 18.37 &  1.03 & 0.2000 & 1991A\&A...246L..24M \\
PQSJ022232.00+330622.1 & QSO B0219+3252 &  144.0787 &  -25.9975 & 17.41 &  1.19 & 0.5040 & 1993AJ....105..660W \\
PQSJ022814.48+311841.3 & NGC   931 &  146.1141 &  -27.1659 & 16.69 &  1.12 & 0.0168 & 2018ApJ...861...49H \\
PQSJ031304.24+461120.7 &  &  146.9644 &   -9.9689 & 19.75 &  0.91 &  &  \\
PQSJ031409.36+471111.0 &  &  146.5947 &   -9.0222 & 18.82 &  1.30 &  &  \\
PQSJ032839.48+451440.5 & 2MASS J03283948+4514404 &  149.7976 &   -9.2689 & 20.14 &  1.05 &  &  \\
PQSJ040821.86+604649.4 & ATO J062.0910+60.7803 &  144.7800 &    6.5612 & 19.74 &  0.72 &  &  \\
PQSJ042630.15+070530.4 & [VV2000] J042630.1+070530 &  187.6921 &  -27.7578 & 17.42 &  1.16 & 0.1700 & 1999A\&AS..134..483H \\
PQSJ042704.25+071632.5 & [OKM2018] SWIFT J0427.0+0734 &  187.6167 &  -27.5359 & 17.37 &  1.16 & 0.0960 & 1990AJ....100.1274W \\
PQSJ045640.88+482058.0 & ATO J074.1703+48.3493 &  158.6602 &    3.2549 & 16.93 &  0.71 &  &  \\
PQSJ052113.09+802100.9 & 2MASS J05211320+8021011 &  132.6207 &   23.0984 & 17.23 &  1.07 &  &  \\
PQSJ052409.90-235455.6 &  &  226.5674 &  -29.1585 & 17.97 &  0.97 &  &  \\
PQSJ053715.36+345344.8 & Gaia DR2 3449698082675788928 &  174.0027 &    1.6442 & 18.29 &  0.55 &  &  \\
PQSJ054323.09-285841.2 & 2MASS J05432311-2858410 &  233.5341 &  -26.6613 & 17.81 &  0.92 & 0.2286 & 2009MNRAS.399..683J \\
PQSJ054324.55-403046.8 &  &  246.1879 &  -29.6119 & 18.62 &  1.10 &  &  \\
PQSJ054508.41-342205.8 & USNO-A2.0 0525-02348261 &  239.4666 &  -27.8375 & 18.71 &  1.18 &  &  \\
PQSJ054917.25-254810.0 & 2MASS J05491725-2548098 &  230.6680 &  -24.4004 & 17.02 &  1.10 & 0.5960 & 2016AJ....152...25M \\
PQSJ055002.09-281312.2 & 2MASS J05500211-2813124 &  233.2415 &  -25.0378 & 17.51 &  0.98 & 0.1520 & 2016AJ....152...25M \\
PQSJ055425.29-280544.8 & UVQS J055425.28-280544.6 &  233.4567 &  -24.0828 & 16.65 &  0.78 & 0.0630 & 2016AJ....152...25M \\
PQSJ055447.23+662043.4 & 2MASS J05544724+6620438 &  147.2725 &   19.2873 & 16.05 &  1.07 & 0.1870 & 2016AJ....152...25M \\
PQSJ055714.69-535714.1 & IRAS 05562-5357 &  261.8740 &  -29.4305 & 17.24 &  1.05 & 0.0967 & 2001ApJ...546..744N \\
PQSJ055925.75-723844.7 &  &  283.3914 &  -29.6666 & 16.66 &  1.00 &  &  \\
PQSJ060105.70-261108.9 & LEDA 3081156 &  232.0496 &  -22.0393 & 16.47 &  0.67 & 0.0390 & 2009MNRAS.399..683J \\
PQSJ060323.15-680437.5 &  &  278.1284 &  -29.4833 & 17.30 &  0.92 &  &  \\
PQSJ060410.06-700422.6 & MGPN LMC 85 &  280.4192 &  -29.3934 & 18.40 &  1.07 &  &  \\
PQSJ060529.04-625450.3 & 1RXS J060529.1-625442 &  272.2162 &  -29.1229 & 17.55 &  1.17 &  &  \\
PQSJ060819.91-715738.1 & IRAS 06091-7157 &  282.5740 &  -29.0191 & 15.64 &  2.04 &  &  \\
PQSJ062204.09-773101.7 & 6dFGS gJ062204.2-773101 &  288.8777 &  -28.0206 & 16.99 &  0.94 & 0.1125 & 2009MNRAS.399..683J \\
PQSJ062255.49-720740.8 & SMP LMC 100 &  282.7771 &  -27.8963 & 18.06 &  1.07 &  &  \\
PQSJ062409.79+273957.6 & 2MASS J06240980+2739574 &  185.2582 &    6.7846 & 19.51 &  0.99 &  &  \\
PQSJ063002.55+690504.3 & QSO B0624+6907 &  145.7085 &   23.3485 & 15.45 &  1.09 & 0.3740 & 2016ApJ...818..113N \\
PQSJ063117.46-290730.7 & ATO J097.8228-29.1251 &  237.5212 &  -16.8695 & 18.45 &  0.77 &  &  \\
PQSJ063224.57+045937.6 & NGC  2244   245 &  206.3163 &   -1.9389 & 18.62 &  0.73 &  &  \\
PQSJ063301.43+622543.2 & 2MASS J06330145+6225436 &  152.7234 &   21.7860 & 17.58 &  1.10 &  &  \\
PQSJ063403.59-744637.8 & 2MASS J06340353-7446377 &  285.8061 &  -27.2379 & 16.35 &  0.97 & 0.1120 & 2014A\&A...561A..67P \\
PQSJ063626.08-684821.2 & [VV2000] J063625.6-684822 &  279.1595 &  -26.4995 & 17.34 &  1.16 & 0.3330 & 2016AJ....152...25M \\
PQSJ063737.12+074819.4 & Gaia DR2 3133893569808649856 &  204.4153 &    0.5038 & 18.73 &  0.64 &  &  \\
PQSJ064011.77-255342.1 & ESO 490-26 &  235.2352 &  -13.7918 & 17.23 &  0.75 & 0.0258 & 2009MNRAS.399..683J \\
PQSJ064134.44-050234.8 & PN We  1-5 &  216.3069 &   -4.4866 & 18.60 &  0.00 &  &  \\
PQSJ064139.39+451002.0 & 2MASS J06413937+4510020 &  170.4583 &   17.2534 & 17.11 &  1.17 & 0.2970 & 2016AJ....152...25M \\
PQSJ064754.31-132823.5 & 2MASS J06475433-1328235 &  224.5843 &   -6.8679 & 18.94 &  0.94 &  &  \\
PQSJ064958.63-073852.7 & 2MASS J06495854-0738522 &  219.5829 &   -3.8019 & 17.43 &  0.74 &  &  \\
PQSJ065030.88-195929.0 & 2MASX J06503094-1959267 &  230.7776 &   -9.1639 & 18.59 &  0.59 & 0.0258 & 2002LEDA.........0P \\
PQSJ065154.68-405846.8 & 2MASS J06515469-4058473 &  250.6734 &  -17.3160 & 18.90 &  1.04 &  &  \\
PQSJ065342.55+165317.8 &  &  198.0526 &    8.1003 & 18.92 &  1.03 &  &  \\
PQSJ065435.35+172204.0 &  &  197.7097 &    8.4991 & 19.07 &  1.08 &  &  \\
PQSJ065629.87-653337.1 & FRL  265 &  276.0061 &  -24.0163 & 15.95 &  0.79 & 0.0305 & 2011MNRAS.416.2840L \\
PQSJ065711.42-105729.1 & 2MASS J06571143-1057290 &  223.3479 &   -3.7124 & 16.44 &  0.99 &  &  \\
PQSJ065746.22-014940.4 & 2MASS J06574621-0149400 &  215.2807 &    0.5748 & 18.97 &  0.88 &  &  \\
PQSJ065814.61+104757.8 &  &  204.0488 &    6.3958 & 19.06 &  0.15 &  &  \\
PQSJ065916.36+123132.1 &  &  202.6035 &    7.3880 & 18.84 &  1.09 &  &  \\
PQSJ070000.47+152252.3 &  &  200.0913 &    8.8024 & 18.06 &  1.39 &  &  \\
PQSJ070001.43+170922.3 & ICRF J070001.5+170921 &  198.4730 &    9.5761 & 19.07 &  1.12 & 1.0800 & 2017ApJS..233....3T \\
PQSJ070139.60+083017.2 &  &  206.4924 &    6.1242 & 18.24 &  0.86 &  &  \\
PQSJ070240.89+114104.1 & LAMOST J070240.89+114104.0 &  203.7361 &    7.7637 & 17.89 &  0.96 &  &  \\
PQSJ070302.13+155735.6 &  &  199.8836 &    9.7101 & 18.52 &  1.15 &  &  \\
PQSJ070331.29-252317.8 &  &  237.0179 &   -8.8083 & 19.35 &  1.37 &  &  \\
PQSJ070344.49+510042.8 & 2MASS J07034450+5100423 &  165.8196 &   22.6111 & 18.04 &  1.12 &  &  \\
PQSJ070418.57-065641.6 & 2MASS J07041857-0656417 &  220.5763 &   -0.3176 & 17.73 &  0.81 &  &  \\
PQSJ070518.98+061219.8 &  &  208.9676 &    5.9025 & 16.89 &  0.78 &  &  \\
PQSJ070529.39+633333.6 & LEDA  138983 &  152.4659 &   25.6308 & 16.15 &  1.03 & 0.1530 & 1991ApJS...76..813S \\
PQSJ070713.20+643600.4 & LEDA   20174 &  151.3597 &   25.9876 & 16.15 &  0.92 & 0.0795 & 2000ApJ...543..552S \\
PQSJ070836.02+131053.0 &  &  203.0160 &    9.7187 & 18.87 &  1.03 &  &  \\
PQSJ070903.38+093508.5 &  &  206.3365 &    8.2405 & 18.46 &  1.12 &  &  \\
PQSJ070919.27+254923.5 & UVQS J070919.25+254923.4 &  191.3114 &   15.1338 & 16.82 &  0.88 & 0.1360 & 2016AJ....152...25M \\
PQSJ071113.13+544653.8 & 2MASS J07111316+5446533 &  162.1339 &   24.6306 & 17.63 &  0.99 &  &  \\
PQSJ071134.79+052347.0 &  &  210.4000 &    6.9299 & 19.26 &  1.31 &  &  \\
PQSJ071139.40-023737.5 & 2MASS J07113941-0237376 &  217.5815 &    3.2924 & 18.67 &  0.92 &  &  \\
PQSJ071156.40+070836.5 &  &  208.8658 &    7.7932 & 18.83 &  1.21 &  &  \\
PQSJ071216.93-832815.0 &  &  295.6722 &  -26.3488 & 18.39 &  1.31 &  &  \\
PQSJ071410.05+063322.3 &  &  209.6457 &    8.0252 & 18.73 &  1.12 &  &  \\
PQSJ071415.08+454156.3 & Mrk  376 &  171.8771 &   22.8371 & 16.71 &  1.06 & 0.0560 & 2016AJ....152...25M \\
PQSJ071507.96+064235.2 & 1RXS J071508.3+064240 &  209.6152 &    8.3083 & 19.10 &  0.92 &  &  \\
PQSJ071708.06+004836.7 &  &  215.1468 &    6.0858 & 19.36 &  1.07 &  &  \\
PQSJ071800.67+440527.6 & ATO J109.5025+44.0908 &  173.7551 &   23.0308 & 15.92 &  0.97 & 0.0634 & 2011MNRAS.416.2840L \\
PQSJ071825.04+194107.6 &  &  197.9989 &   14.5971 & 17.81 &  1.08 &  &  \\
PQSJ071831.03+084248.2 &  &  208.1743 &    9.9495 & 18.83 &  1.31 &  &  \\
PQSJ071938.07+542637.6 & 2MASS J07193801+5426372 &  162.7791 &   25.7520 & 17.73 &  0.94 &  &  \\
PQSJ071945.14+315347.3 & ATO J109.9381+31.8962 &  186.2971 &   19.5148 & 17.42 &  0.97 &  &  \\
PQSJ071950.85+742757.0 & QSO B0713+745 &  140.4417 &   27.9564 & 16.91 &  1.20 & 0.4750 & 2016ApJ...818..113N \\
PQSJ072001.62+181726.3 & EM* StHA   62 &  199.4742 &   14.3736 & 14.19 &  0.96 &  &  \\
PQSJ072049.96-135134.1 & 2MASS J07204996-1351340 &  228.5853 &    0.0742 & 18.82 &  1.08 &  &  \\
PQSJ072051.14-021033.9 & 2MASS J07205117-0210335 &  218.2421 &    5.5386 & 18.01 &  0.90 &  &  \\
PQSJ072108.99-251033.6 & IRAS 07191-2504 &  238.6349 &   -5.1548 & 18.03 &  0.56 &  &  \\
PQSJ072153.36+712036.2 & 8C 0716+714 &  143.9811 &   28.0176 & 15.10 &  1.00 & 0.3000 & 2016ApJ...818..113N \\
PQSJ072217.53+303050.5 & LEDA 1906770 &  187.8728 &   19.5384 & 16.86 &  1.01 & 0.1000 & 2016AJ....152...25M \\
PQSJ072241.80+023436.8 & 2MASS J07224180+0234369 &  214.2009 &    8.1284 & 17.35 &  0.95 &  &  \\
PQSJ072331.84-014626.7 & 2MASS J07233186-0146267 &  218.1947 &    6.3178 & 18.53 &  1.15 &  &  \\
PQSJ072434.03-000558.5 &  &  216.8187 &    7.3200 & 18.39 &  0.94 &  &  \\
PQSJ072605.44-012730.9 &  &  218.2111 &    7.0310 & 19.02 &  1.13 &  &  \\
PQSJ072743.77-001418.2 &  &  217.3103 &    7.9579 & 17.60 &  1.22 &  &  \\
PQSJ072839.16+015905.6 &  &  215.4196 &    9.1831 & 19.13 &  1.33 &  &  \\
PQSJ073309.17+455506.5 & 2MASS J07330920+4555062 &  172.5508 &   26.0804 & 16.89 &  0.95 & 0.1414 & 2009ApJS..182..543A \\
PQSJ073359.63-041814.0 &  &  221.6760 &    7.4481 & 18.83 &  0.98 &  &  \\
PQSJ073405.87+135316.7 & UVQS J073405.85+135316.7 &  205.0720 &   15.6250 & 17.11 &  1.06 & 0.3060 & 2016AJ....152...25M \\
PQSJ073623.17+392618.1 & 2MASX J07362309+3926173 &  179.6768 &   25.0625 & 16.60 &  0.97 & 0.1181 & 2009ApJS..182..543A \\
PQSJ073710.03-131133.7 &  &  229.8967 &    3.8809 & 18.57 &  1.03 &  &  \\
PQSJ073807.05+212730.6 & 2MASS J07380702+2127295 &  198.1950 &   19.5333 & 16.76 &  1.33 & 0.7520 & 2016AJ....152...25M \\
PQSJ073833.88-020423.1 & 3C 185 &  220.2272 &    9.5035 & 18.49 &  1.26 & 1.0330 & 2002LEDA.........0P \\
PQSJ073949.44+532314.4 &  &  164.5076 &   28.5057 & 17.47 &  1.15 &  &  \\
PQSJ074126.40+354703.0 & 2MASS J07412637+3547027 &  183.8584 &   25.0180 & 17.78 &  0.95 & 0.1328 & 2012AJ....143...64J \\
PQSJ074227.08+465643.0 & LEDA 3095768 &  171.7961 &   27.8604 & 18.03 &  0.94 & 0.1678 & 2012ApJS..203...21A \\
PQSJ074232.82+494834.2 & Mrk   79 &  168.6027 &   28.3822 & 15.67 &  1.08 & 0.0223 & 2015A\&A...573A..59G \\
PQSJ074312.55+742936.8 & 2E  1852 &  140.3544 &   29.5169 & 17.33 &  0.99 & 0.3120 & 1991ApJS...76..813S \\
PQSJ074522.77-135543.3 &  &  231.5213 &    5.2602 & 17.20 &  0.61 &  &  \\
PQSJ074541.63+314256.7 & ICRF J074541.6+314256 &  188.4296 &   24.6626 & 16.54 &  1.06 & 0.4609 & 2012ApJS..203...21A \\
PQSJ074552.29-080816.7 &  &  226.4996 &    8.2067 & 18.25 &  1.14 &  &  \\
PQSJ074706.93-113421.0 &  &  229.6714 &    6.7921 & 18.63 &  1.17 &  &  \\
PQSJ074906.53+451034.6 & 7C 074532.29+451808.00 &  174.0155 &   28.6613 & 17.04 &  1.02 & 0.1922 & 2018A\&A...613A..51P \\
PQSJ074910.59+284214.5 & 2MASS J07491060+2842145 &  191.8329 &   24.4308 & 17.56 &  0.99 & 0.3440 & 2012ApJS..203...21A \\
PQSJ074948.30+345444.3 & 2MASS J07494826+3454439 &  185.3158 &   26.4127 & 17.25 &  0.94 & 0.1323 & 2012ApJS..203...21A \\
PQSJ074950.82+335743.9 & IC 2207 &  186.3365 &   26.1569 & 17.63 &  0.75 & 0.0160 & 2015ApJS..219...12A \\
PQSJ075004.42+170244.7 & 2MASS J07500439+1702459 &  203.6865 &   20.4260 & 17.38 &  1.07 & 0.3683 & 2012ApJS..203...21A \\
PQSJ075051.06+775101.6 &  &  136.4612 &   29.5899 & 17.13 &  1.23 &  &  \\
PQSJ075100.75+032041.0 & 2MASX J07510082+0320401 &  216.7908 &   14.7715 & 15.72 &  1.00 & 0.0990 & 2002LEDA.........0P \\
PQSJ075112.30+291938.9 & QSO B0748+295 &  191.3377 &   25.0497 & 16.30 &  1.28 & 0.9142 & 2012ApJS..203...21A \\
PQSJ075126.22+360949.3 & 2MASS J07512622+3609492 &  184.0651 &   27.0674 & 17.92 &  0.95 & 0.1482 & 2012ApJS..203...21A \\
PQSJ075225.74+283040.0 & 2MASS J07522573+2830399 &  192.2860 &   25.0451 & 17.50 &  0.61 & 0.0618 & 2012ApJS..203...21A \\
PQSJ075441.42+010149.7 &  &  219.3511 &   14.5180 & 17.63 &  1.19 &  &  \\
PQSJ075459.48-041918.9 & 2MASS J07545948-0419190 &  224.2267 &   12.0398 & 17.83 &  1.01 &  &  \\
PQSJ075525.29+391109.7 & [OKM2018] SWIFT J0755.7+3912 &  180.9678 &   28.5683 & 16.09 &  0.69 & 0.0332 & 2012A\&A...545A..15S \\
PQSJ075649.44-095349.9 & SDSS J075649.43-095349.8 &  229.3978 &    9.6937 & 19.11 &  1.13 &  &  \\
PQSJ075706.65+095635.5 & QSO B0754+10 &  211.3115 &   19.0571 & 17.10 &  1.10 & 0.2660 & 2011NewA...16..503M \\
PQSJ075730.37+152453.2 & ATO J119.3765+15.4147 &  206.0476 &   21.4154 & 17.17 &  1.04 & 0.3886 & 2012ApJS..203...21A \\
PQSJ075800.07+392029.3 & 1E 0754+39.3 &  180.9215 &   29.0896 & 16.03 &  0.95 & 0.0950 & 2016AJ....152...25M \\
PQSJ075819.70+421934.7 & LEDA 2196699 &  177.5812 &   29.7600 & 17.11 &  1.05 & 0.2112 & 2018A\&A...613A..51P \\
PQSJ080136.82-063609.4 & 2MASS J08013683-0636095 &  227.0867 &   12.3608 & 17.65 &  0.75 & 0.0862 & 2009MNRAS.399..683J \\
PQSJ080405.93+050649.5 & RC J0804+0506 &  216.6961 &   18.4817 & 16.93 &  1.39 &  &  \\
PQSJ080452.73+212049.6 & LEDA  139051 &  200.7895 &   25.2831 & 16.86 &  0.99 & 0.1244 & 2012ApJS..203...21A \\
PQSJ080554.10+111432.8 & 2MASS J08055411+1114324 &  211.0375 &   21.5648 & 17.52 &  1.05 & 0.2274 & 2012ApJS..203...21A \\
PQSJ080627.14-211621.7 &  &  240.4401 &    5.8083 & 18.90 &  1.23 &  &  \\
PQSJ080634.87-283159.8 & WRAY 15-157 &  246.6045 &    1.9475 & 18.95 &  1.07 & 0.0003 & 2018yCat.1345....0G \\
PQSJ080919.34-202653.8 &  &  240.0980 &    6.8151 & 18.32 &  1.19 &  &  \\
PQSJ080954.39+074355.6 & 2MASS J08095437+0743553 &  214.8929 &   20.9426 & 17.06 &  1.11 & 0.6521 & 2012ApJS..203...21A \\
PQSJ081028.70-165220.6 &  &  237.1895 &    8.9512 & 19.14 &  1.36 &  &  \\
PQSJ081119.31-180252.8 &  &  238.3050 &    8.4963 & 19.19 &  1.27 &  &  \\
PQSJ081303.50-205251.9 &  &  240.9393 &    7.3190 & 19.00 &  1.44 &  &  \\
PQSJ081459.52-181520.8 &  &  238.9557 &    9.1206 & 19.14 &  1.17 &  &  \\
PQSJ081522.96+015500.4 & ICRF J081522.9+015459 &  221.0706 &   19.5006 & 17.24 &  1.11 & 0.4000 & 2017ApJS..233....3T \\
PQSJ081629.39-180906.8 &  &  239.0618 &    9.4761 & 18.99 &  1.01 &  &  \\
PQSJ081807.57+024932.2 & UVQS J081807.53+024932.1 &  220.5604 &   20.5330 & 17.26 &  0.64 & 0.0630 & 2016AJ....152...25M \\
PQSJ082045.40+130618.7 & [VV2000] J082045.4+130618 &  210.8302 &   25.6404 & 16.63 &  1.35 & 1.1248 & 2009ApJS..182..543A \\
PQSJ082347.03-201252.2 &  &  241.7693 &    9.7837 & 18.89 &  1.51 &  &  \\
PQSJ082559.42-135142.5 & UVQS J082559.37-135142.7 &  236.6484 &   13.7220 & 17.43 &  1.11 & 0.3190 & 2016AJ....152...25M \\
PQSJ082628.22+085535.7 & 2MASS J08262820+0855354 &  215.6789 &   25.1486 & 17.73 &  1.02 & 0.2543 & 2009ApJS..182..543A \\
PQSJ082633.57+074248.3 & QSO J0826+0742 &  216.8905 &   24.6344 & 17.37 &  1.08 & 0.3113 & 2012ApJS..203...21A \\
PQSJ082930.52+081238.3 & LEDA 1341762 &  216.7553 &   25.5093 & 17.33 &  1.04 & 0.1291 & 2009ApJS..182..543A \\
PQSJ083011.01-250252.0 &  &  246.6490 &    8.2618 & 18.78 &  1.29 &  &  \\
PQSJ083016.49-672528.6 & 2MASX J08301655-6725289 &  281.6422 &  -16.1542 & 16.96 &  0.69 & 0.0348 & 2009MNRAS.399..683J \\
PQSJ083029.00+024304.1 & 2MASS J08302893+0243038 &  222.2342 &   23.2014 & 17.74 &  1.00 &  &  \\
PQSJ083356.46-010113.0 & 6dFGS gJ083356.5-010112 &  226.2272 &   22.1314 & 16.49 &  1.20 & 0.2907 & 2009MNRAS.399..683J \\
PQSJ083452.34-233540.1 &  &  246.0759 &    9.9743 & 18.87 &  0.91 &  &  \\
PQSJ083641.49-040808.7 &  &  229.4885 &   21.1372 & 17.57 &  1.20 &  &  \\
PQSJ083752.29-244028.5 &  &  247.3715 &    9.8962 & 18.82 &  1.24 &  &  \\
PQSJ084219.12-034931.4 & 2MASS J08421910-0349314 &  229.9894 &   22.4972 & 17.11 &  1.19 & 0.3566 & 2009MNRAS.399..683J \\
PQSJ084320.25-264017.3 & 2MASS J08432025-2640177 &  249.7434 &    9.6936 & 18.40 &  1.26 &  &  \\
PQSJ084553.53-024100.2 & [VV2006] J084553.4-024100 &  229.4348 &   23.8535 & 16.64 &  1.10 & 0.4680 & 2018AJ....155..189D \\
PQSJ085259.18+031320.4 & 2MASS J08525922+0313207 &  224.7534 &   28.3666 & 17.33 &  1.03 & 0.2968 & 2012ApJS..203...21A \\
PQSJ085506.79-030336.5 & 2MASS J08550679-0303368 &  231.1207 &   25.6137 & 17.67 &  0.98 &  &  \\
PQSJ085915.68+011800.4 & 2MASS J08591566+0118006 &  227.5403 &   28.7681 & 17.72 &  1.03 & 0.2821 & 2012ApJS..203...21A \\
PQSJ090420.87+013038.4 & [VV2006] J090420.9+013038 &  228.0758 &   29.9692 & 17.19 &  1.27 & 0.7921 & 2012ApJS..203...21A \\
PQSJ090553.67-042612.3 & 2MASS J09055360-0426119 &  234.0405 &   27.1257 & 16.46 &  1.05 & 0.3700 & 2016AJ....152...25M \\
PQSJ091134.96-134801.4 & 2MASS J09113489-1348022 &  243.3322 &   22.7939 & 17.93 &  1.07 &  &  \\
PQSJ091945.29-063225.5 & 2MASS J09194530-0632252 &  238.2226 &   28.7612 & 16.65 &  1.02 & 0.7061 & 2009MNRAS.399..683J \\
PQSJ092100.39-051545.7 & 2MASS J09210036-0515465 &  237.2411 &   29.7594 & 17.36 &  1.06 & 0.3800 & 2016AJ....152...25M \\
PQSJ092101.13-134252.0 &  &  244.8115 &   24.6442 & 18.82 &  1.07 &  &  \\
PQSJ092233.99-124225.6 &  &  244.2060 &   25.5694 & 18.55 &  1.14 &  &  \\
PQSJ092315.05-132826.5 & 1RXS J092315.2-132817 &  244.9822 &   25.2169 & 18.18 &  0.76 &  &  \\
PQSJ092443.26-130344.2 &  &  244.8787 &   25.7543 & 18.76 &  1.36 &  &  \\
PQSJ092751.79-203451.5 & ICRF J092751.8-203451 &  251.6378 &   21.4229 & 16.76 &  1.04 & 0.3474 & 2009MNRAS.399..683J \\
PQSJ092815.22-122132.6 &  &  244.8797 &   26.8664 & 18.85 &  1.13 &  &  \\
PQSJ093318.16-171440.3 & [VV2006] J093318.1-171441 &  249.8788 &   24.6114 & 17.66 &  1.19 & 0.3100 & 2000ApJS..129..547B \\
PQSJ093622.14-113433.9 & 2MASS J09362212-1134342 &  245.6347 &   28.8930 & 17.78 &  0.75 & 0.0912 & 2009MNRAS.399..683J \\
PQSJ093846.25-215547.5 & 2MASS J09384620-2155477 &  254.5713 &   22.3549 & 15.73 &  1.06 & 0.2487 & 2009MNRAS.399..683J \\
PQSJ093911.30-111145.7 & 2MASS J09391128-1111454 &  245.8140 &   29.6659 & 17.59 &  0.99 &  &  \\
PQSJ094745.72-215547.6 & [VV96] J094745.8-215547 &  256.1721 &   23.8313 & 17.53 &  1.24 & 0.3160 & 2016AJ....152...25M \\
PQSJ095513.58-212303.1 & LEDA  828986 &  257.1338 &   25.4255 & 17.22 &  0.67 & 0.1085 & 2009MNRAS.399..683J \\
PQSJ095912.22-185250.1 & EC 09568-1838 &  255.9711 &   27.8872 & 16.06 &  1.23 & 0.5720 & 2016AJ....152...25M \\
PQSJ100159.80-443802.0 & QSO B0959-443 &  273.9334 &    8.4888 & 17.77 &  1.24 & 0.8370 & 1989ApJS...69....1H \\
PQSJ100826.52-273544.3 & 2MASS J10082651-2735443 &  264.0898 &   22.7249 & 17.00 &  1.06 & 0.3080 & 2016AJ....152...25M \\
PQSJ102239.94-302930.6 & 2MASS J10223994-3029305 &  268.6664 &   22.3055 & 17.40 &  1.07 & 0.3165 & 2009MNRAS.399..683J \\
PQSJ102542.52-265349.9 & 1RXS J102542.9-265354 &  266.9630 &   25.6138 & 17.74 &  0.93 &  &  \\
PQSJ103126.83-273959.2 &  &  268.6109 &   25.7201 & 18.34 &  1.11 &  &  \\
PQSJ103236.07-281326.9 & QSO B1030-2757 &  269.1928 &   25.4032 & 16.37 &  1.06 & 0.1484 & 2009MNRAS.399..683J \\
PQSJ104833.80-390237.8 & 2MASS J10483382-3902379 &  278.2949 &   17.8787 & 16.82 &  0.88 & 0.0450 & 2016AJ....152...25M \\
PQSJ105251.04-272716.1 & HE 1050-2711 &  272.9878 &   28.4336 & 16.49 &  0.99 & 0.2070 & 2016AJ....152...25M \\
PQSJ105727.86-403940.5 & 6dFGS gJ105727.9-403941 &  280.6670 &   17.2315 & 17.35 &  1.03 & 0.3981 & 2009MNRAS.399..683J \\
PQSJ110106.88-303605.7 & EC 10587-3019 &  276.4788 &   26.5423 & 16.73 &  1.07 & 0.3257 & 2009MNRAS.399..683J \\
PQSJ110331.56-325116.7 & CTS   10 &  278.1173 &   24.7652 & 16.91 &  1.10 & 0.3555 & 2009MNRAS.399..683J \\
PQSJ111028.89-603155.9 & 2MASS J11102887-6031558 &  290.8237 &   -0.0558 & 18.24 &  0.61 & 0.0000 & 2020AJ....160..120J \\
PQSJ111704.03-290230.5 & CTS   11 &  279.2387 &   29.4519 & 16.88 &  1.03 & 0.0700 & 2016AJ....152...25M \\
PQSJ112048.59-293939.3 & EC 11183-2923 &  280.3805 &   29.2117 & 17.34 &  1.25 &  &  \\
PQSJ112731.83-304445.6 & HE 1125-3028 &  282.3863 &   28.7563 & 16.71 &  1.27 & 0.6680 & 2016AJ....152...25M \\
PQSJ112855.15-343214.0 &  &  284.2046 &   25.3158 & 17.75 &  0.97 &  &  \\
PQSJ112914.99-383730.3 & 2MASS J11291493-3837306 &  285.7926 &   21.4958 & 17.00 &  0.96 & 0.2328 & 2009MNRAS.399..683J \\
PQSJ114755.66-385809.3 &  &  289.6448 &   22.2583 & 17.40 &  1.13 &  &  \\
PQSJ115227.68-374837.7 & 2MASS J11522769-3748378 &  290.2564 &   23.6053 & 17.80 &  1.05 &  &  \\
PQSJ121514.39-322101.8 & 2MASS J12151434-3221016 &  294.1150 &   29.9007 & 17.68 &  0.99 &  &  \\
PQSJ123137.19-475802.5 & 2MASX J12313717-4758019 &  299.5034 &   14.7724 & 16.41 &  0.82 & 0.0279 & 2009MNRAS.399..683J \\
PQSJ130421.04-435310.5 & 2MASS J13042100-4353102 &  305.3909 &   18.9238 & 16.91 &  0.86 &  &  \\
PQSJ130809.02-455417.0 & 2MASS J13080899-4554180 &  305.9687 &   16.8685 & 17.28 &  1.05 & 0.3320 & 2016AJ....152...25M \\
PQSJ132225.68-325431.3 &  &  310.4045 &   29.5130 & 16.25 &  0.62 &  &  \\
PQSJ133325.99-340052.8 & ESO 383-18 &  312.7862 &   28.0522 & 17.06 &  1.31 & 0.0130 & 2004MNRAS.355..273C \\
PQSJ140806.74-302354.3 & 2MAXI J1406-300 &  321.9462 &   29.6266 & 16.53 &  1.04 & 0.0250 & 2016AJ....152...25M \\
PQSJ152052.55-213358.3 & 2MASS J15205249-2133581 &  343.2845 &   29.3584 & 17.45 &  1.06 & 0.1502 & 2009MNRAS.399..683J \\
PQSJ152546.60-235430.2 &  &  342.6406 &   26.8207 & 18.61 &  1.06 &  &  \\
PQSJ160551.09-611144.5 & LEDA 3079933 &  324.2714 &   -6.6615 & 18.79 &  0.81 & 0.0520 & 2008A\&A...482..113M \\
PQSJ161836.16-592716.5 & LEDA 3078020 &  326.6297 &   -6.4847 & 18.76 &  0.69 & 0.0356 & 2011MNRAS.416.2840L \\
PQSJ162526.51+852941.9 & LEDA   58086 &  118.7615 &   29.7461 & 17.47 &  0.85 & 0.0629 & 2002LEDA.........0P \\
PQSJ164533.35+022059.6 & UVQS J164533.31+022059.3 &   19.6376 &   28.8772 & 16.83 &  1.15 & 0.6450 & 2016AJ....152...25M \\
PQSJ165605.59-520340.3 & 2MASS J16560561-5203408 &  335.6882 &   -5.4921 & 18.52 &  0.98 & 0.0540 & 2006A\&A...459...21M \\
PQSJ170025.18-724044.7 & 2MASS J17002537-7240450 &  319.0313 &  -18.1745 & 17.65 &  0.86 & 0.1047 & 2009MNRAS.399..683J \\
PQSJ172421.73+063837.9 & 2MASX J17242174+0638371 &   28.9010 &   22.3712 & 16.97 &  0.79 & 0.0220 & 2016AJ....152...25M \\
PQSJ172422.82-513410.1 &  &  338.6679 &   -8.7477 & 18.83 &  1.38 &  &  \\
PQSJ173840.53+322409.2 & 7C 173649.10+322532.00 &   56.9083 &   28.6152 & 16.90 &  0.91 & 0.1300 & 2017ApJS..233....3T \\
PQSJ174128.11+034852.0 & IRAS 17389+0350 &   28.2732 &   17.2928 & 17.53 &  1.17 & 0.0230 & 2016AJ....152...25M \\
PQSJ174206.89+182721.1 & 4C 18.51 &   42.5163 &   23.3181 & 17.24 &  1.00 & 0.1860 & 2016ApJ...818..113N \\
PQSJ174506.56-020843.7 &  &   23.2521 &   13.6946 & 18.02 &  1.04 &  &  \\
PQSJ174638.78+411047.3 & 2MASS J17463884+4110467 &   67.0808 &   29.1495 & 17.27 &  1.01 &  &  \\
PQSJ174704.02+260336.2 & LAMOST J174704.02+260337.0 &   50.7509 &   24.9571 & 18.43 &  1.02 & 0.3403 & 2018AJ....155..189D \\
PQSJ174707.17+383519.4 &  &   64.2172 &   28.4954 & 17.71 &  1.03 &  &  \\
PQSJ175044.52+255518.0 & SDSS J175044.50+255518.1 &   50.9117 &   24.1326 & 18.47 &  1.03 & 0.3900 & 2018AJ....155..189D \\
PQSJ175114.15+253931.5 &  &   50.6821 &   23.9389 & 18.94 &  1.30 &  &  \\
PQSJ180055.04+423957.1 & 2MASS J18005509+4239572 &   69.3756 &   26.8802 & 17.14 &  0.99 &  &  \\
PQSJ180249.59+512007.5 &  &   79.0442 &   28.2356 & 17.45 &  1.02 &  &  \\
PQSJ180650.80+694928.4 & 3C 371 &  100.1314 &   29.1668 & 17.07 &  0.86 & 0.0510 & 2016AJ....152...25M \\
PQSJ181234.59+395126.5 & 2MASS J18123455+3951260 &   66.9813 &   24.0494 & 17.25 &  1.01 &  &  \\
PQSJ181828.96+674125.6 & 2E  4057 &   97.7368 &   28.0517 & 16.45 &  1.03 & 0.3140 & 2004ApJ...617..192M \\
PQSJ182211.47-700219.2 & UVQS J182211.57-700219.5 &  324.5067 &  -22.9496 & 17.84 &  1.17 & 0.3530 & 2016AJ....152...25M \\
PQSJ182408.93+464017.1 & LEDA  165770 &   74.7816 &   23.8437 & 16.95 &  0.83 & 0.0696 & 2011MNRAS.416.2840L \\
PQSJ182446.44+650924.7 & HS 1824+6507 &   94.9492 &   27.2062 & 17.21 &  1.11 & 0.3000 & 1995A\&A...300..323J \\
PQSJ182754.15+095853.6 & LS  IV +09    4 &   39.2359 &    9.7234 & 14.83 &  0.74 &  &  \\
PQSJ183023.19+731310.6 & 2MASS J18302317+7313107 &  104.0389 &   27.3960 & 14.67 &  0.96 & 0.1230 & 2001AJ....121.2843B \\
PQSJ183046.48+094459.2 &  &   39.3408 &    8.9862 & 19.79 &  0.89 &  &  \\
PQSJ183249.71+534021.8 & LEDA 2447627 &   82.5863 &   24.2128 & 17.12 &  0.64 & 0.0390 & 2016ApJ...818..113N \\
PQSJ183503.36+324146.5 & ICRF J183503.3+324146 &   61.3051 &   17.4461 & 15.71 &  0.94 & 0.0581 & 2002LEDA.........0P \\
PQSJ183849.18+480234.4 & ATO J279.7048+48.0428 &   76.9498 &   21.8289 & 17.21 &  0.73 & 0.3000 & 2013MNRAS.430.2464M \\
PQSJ184640.81+635128.7 & UVQS J184640.72+635128.5 &   93.9346 &   24.6757 & 17.25 &  1.19 & 0.4080 & 2016AJ....152...25M \\
PQSJ185700.07+160626.6 & 2MASS J18570009+1606261 &   47.9428 &    6.0707 & 19.39 &  0.77 &  &  \\
PQSJ185801.14+485023.3 & 1RXS J185800.9+485020 &   78.8222 &   19.0554 & 17.69 &  0.86 & 0.0790 & 2012ApJ...751...52E \\
PQSJ190525.86+422739.2 & Z 229-15 &   73.0896 &   15.5315 & 17.76 &  0.70 & 0.0275 & 2002LEDA.........0P \\
PQSJ190938.94-622854.5 & ESO 104-41 &  333.7333 &  -25.9149 & 16.70 &  0.96 & 0.0852 & 2009MNRAS.399..683J \\
PQSJ191928.08-295808.3 & LEDA 2829969 &    8.1819 &  -18.7741 & 16.00 &  1.05 & 0.1668 & 2009MNRAS.399..683J \\
PQSJ192114.11-584013.4 & ESO 141-55 &  338.1833 &  -26.7111 & 14.61 &  0.87 & 0.0371 & 2009MNRAS.399..683J \\
PQSJ192447.35-512427.3 & UVQS J192447.29-512426.4 &  346.2566 &  -25.9062 & 16.44 &  1.20 & 0.7660 & 2016AJ....152...25M \\
PQSJ192933.39-390841.9 & UVQS J192933.41-390842.1 &  359.5861 &  -23.7113 & 17.62 &  1.08 & 0.4230 & 2016AJ....152...25M \\
PQSJ193008.30+020449.4 &  &   39.2395 &   -7.6383 & 19.36 &  1.33 &  &  \\
PQSJ193423.21-515129.4 & UVQS J193423.24-515129.8 &  346.0735 &  -27.4535 & 18.06 &  1.14 & 0.1690 & 2016AJ....152...25M \\
PQSJ193804.21-510949.3 & 2MASS J19380439-5109493 &  346.9551 &  -27.9020 & 16.32 &  0.70 & 0.0400 & 2009MNRAS.399..683J \\
PQSJ193810.75+540856.0 &  &   86.4448 &   15.3090 & 17.55 &  1.10 &  &  \\
PQSJ193819.64-432645.9 & 2MASS J19381959-4326461 &  355.4683 &  -26.4134 & 16.28 &  1.02 & 0.0791 & 2009MNRAS.399..683J \\
PQSJ193929.56+700748.8 & 2MASS J19392938+7007490 &  102.0032 &   21.4616 & 17.30 &  1.00 & 0.1200 & 2016ApJ...818..113N \\
PQSJ194357.73-140911.5 & NGC  6818 &   25.8586 &  -17.9112 & 12.92 &  1.87 & -0.0000 & 1953GCRV..C......0W \\
PQSJ194545.28-474735.1 &  &  350.9649 &  -28.5639 & 17.71 &  1.01 &  &  \\
PQSJ194635.31+041727.5 &  &   43.1722 &  -10.2178 & 17.48 &  0.71 &  &  \\
PQSJ194658.72+052021.8 & 1RXS J194658.6+052031 &   44.1538 &   -9.8004 & 19.88 &  1.10 &  &  \\
PQSJ194732.38-421750.0 & UVQS J194732.37-421750.0 &  357.1506 &  -27.8012 & 17.11 &  1.02 & 0.2420 & 2016AJ....152...25M \\
PQSJ195041.28-200021.6 & UVQS J195041.29-200021.6 &   20.8204 &  -21.7034 & 17.09 &  0.97 & 0.2920 & 2016AJ....152...25M \\
PQSJ195904.13+105335.5 &  &   50.5527 &   -9.6715 & 18.86 &  1.32 &  &  \\
PQSJ200631.82-153906.5 & IRAS 20037-1547 &   26.7465 &  -23.5041 & 17.80 &  1.00 & 0.1919 & 2011MNRAS.414..500H \\
PQSJ202202.37-182259.9 & 2MASX J20220235-1822587 &   25.5130 &  -27.9729 & 17.53 &  0.77 & 0.0549 & 2009MNRAS.399..683J \\
PQSJ202610.07+112928.0 & 2MASS J20261004+1129279 &   54.6127 &  -15.0388 & 17.04 &  0.99 &  &  \\
PQSJ203202.53-102330.9 & 2MASS J20320251-1023314 &   34.8669 &  -27.0196 & 17.85 &  0.91 &  &  \\
PQSJ203311.82+173234.1 & UVQS J203311.76+173234.0 &   60.7865 &  -13.1253 & 17.22 &  1.03 & 0.2340 & 2016AJ....152...25M \\
PQSJ203335.87-032038.6 & 2MASS J20333589-0320385 &   42.0361 &  -24.1831 & 17.07 &  1.07 & 0.6960 & 2016AJ....152...25M \\
PQSJ203413.30-040529.0 & LEDA 1062999 &   41.3933 &  -24.6733 & 16.52 &  0.84 &  &  \\
PQSJ204237.21+750802.1 & ICRF J204237.3+750802 &  108.9977 &   19.5267 & 17.23 &  0.99 & 0.1040 & 2002LEDA.........0P \\
PQSJ204303.53-010125.0 & [VV2006] J204303.6-010126 &   45.5299 &  -25.1254 & 18.35 &  1.43 & 1.1899 & 2009ApJS..182..543A \\
PQSJ204306.26+032451.4 & 2MASS J20430626+0324516 &   49.7200 &  -22.8787 & 17.47 &  1.13 & 0.2710 & 1995A\&AS..110..469B \\
PQSJ204409.71-104324.3 & Mrk  509 &   35.9711 &  -29.8550 & 14.50 &  0.89 & 0.0350 & 2016ApJ...818..113N \\
PQSJ204620.87-024845.2 & Mrk  896 &   44.2456 &  -26.7197 & 16.36 &  0.66 & 0.0268 & 2011MNRAS.416.2840L \\
PQSJ205256.03+024546.2 &  &   50.5220 &  -25.3097 & 16.88 &  0.89 &  &  \\
PQSJ205820.58+340138.6 &  &   77.5978 &   -7.6261 & 18.62 &  1.31 &  &  \\
PQSJ210931.91+353258.5 & ICRF J210931.8+353257 &   80.2819 &   -8.3528 & 17.44 &  0.99 & 0.2015 & 2013ApJ...767...14P \\
PQSJ211420.95+252422.9 & ATO J318.5876+25.4062 &   73.2535 &  -15.8905 & 17.50 &  1.13 &  &  \\
PQSJ211452.57+060742.9 & 2MASS J21145258+0607423 &   57.0403 &  -28.0144 & 16.58 &  1.20 & 0.4570 & 2017AJ....154..114O \\
PQSJ211547.08+190843.0 & [VV2000] J211547.1+190843 &   68.4845 &  -20.2226 & 16.79 &  1.20 & 0.4100 & 2016AJ....152...25M \\
PQSJ214153.44+315127.1 & 2MASX J21415350+3151282 &   82.4352 &  -15.7309 & 17.55 &  0.66 & 0.0433 & 2002LEDA.........0P \\
PQSJ214335.53+174349.4 & ICRF J214335.5+174348 &   72.1156 &  -26.0841 & 16.92 &  1.11 & 0.2130 & 2016ApJ...818..113N \\
PQSJ215647.45+224250.2 & QSO J2156+2242 &   78.4262 &  -24.6981 & 15.97 &  1.24 & 1.2900 & 2016ApJ...818..113N \\
PQSJ220314.97+314538.5 & 4C 31.63 &   85.9568 &  -18.7787 & 16.24 &  1.03 & 0.2970 & 2016ApJ...818..113N \\
PQSJ220651.79+275758.6 & 2MASS J22065183+2757584 &   84.0498 &  -22.2169 & 17.24 &  1.01 &  &  \\
PQSJ221536.83+290235.4 & 4C 28.53 &   86.4035 &  -22.5622 & 16.69 &  1.02 & 0.2290 & 1999A\&AS..139..575W \\
PQSJ223954.44+272431.9 & LEDA 1807248 &   90.1932 &  -26.9743 & 17.38 &  0.94 & 0.0654 & 2020ApJS..250....8L \\
PQSJ224436.88+475301.3 & 1RXS J224436.9+475306 &  101.9768 &   -9.7610 & 18.83 &  1.28 &  &  \\
PQSJ225147.70+341930.6 & LEDA 3096705 &   96.4956 &  -22.3319 & 17.42 &  1.02 & 0.1320 & 1999A\&AS..139..575W \\
PQSJ225603.39+273209.7 & HS 2253+2716 &   93.7370 &  -28.7147 & 17.07 &  1.18 & 0.3635 & 2020ApJS..250....8L \\
PQSJ232026.52+592924.3 & IRAS 23182+5912 &  111.6068 &   -1.3658 & 19.23 &  1.01 &  &  \\
PQSJ234229.32+530025.5 & 2MASS J23422932+5300254 &  112.5463 &   -8.4685 & 19.23 &  1.02 &  &  \\
PQSJ235217.32+332329.9 & LEDA 2031036 &  108.9824 &  -27.8727 & 17.55 &  0.93 &  &  \\
\enddata
\tablecomments{{}  Properties of Plane QSO candidates, including their PQS and SIMBAD names, their Galactic coordinates, their GALEX NUV magnitudes, and W1-W2 colors. 165 of these 305 PQS QSO candidates have spectroscopic redshifts available in the literature, and the last two columns of this table list the published redshifts and their sources, when available. }
\end{deluxetable*}

 \startlongtable
\begin{deluxetable*}{lccccccccc}
\tablewidth{0pc}
\tablecaption{Plane QSO Candidate Spectroscopic Redshifts\label{tab:pqsoprops}}
\tabletypesize{\scriptsize}
\tablehead{\colhead{PQS Name}  & \colhead{$l$}  &\colhead{$b$} &\colhead{m$_{\rm NUV}$} & \colhead{Telescope} & \colhead{Type } & \colhead{z$_{\rm narrow}$}  &  \colhead{z$_{\rm MARZ}$} & \colhead{Ref z} &\colhead{Ref} } 
\startdata
PQSJ002324.76+605103.6 &  119.5240 &   -1.8350 & 16.50 & APO-3.5m & Star &   & 0.0000 &  &  \\
PQSJ004342.52+372520.1 &  121.2328 &  -25.4239 & 16.74 & MDM-2.4m & AGN  & 0.0797 & 0.0801 & 0.0800 & 2015RAA....15.1438H \\
PQSJ004818.93+394112.1 &  122.2785 &  -23.1810 & 16.71 & APO-3.5m & AGN  & 0.1349 & 0.1349 & 0.1340 & 2016ApJ...818..113N \\
PQSJ005050.78+353644.0 &  122.7967 &  -27.2593 & 16.47 & MDM-2.4m & AGN  & 0.0581 & 0.0583 & 0.0580 & 2016AJ....152...25M \\
PQSJ010131.18+422935.1 &  124.9137 &  -20.3397 & 16.94 & APO-3.5m & AGN  & 0.1848 & 0.1850 & 0.1900 & 2016ApJ...818..113N \\
PQSJ010415.84+402243.5 &  125.5740 &  -22.4271 & 16.50 & APO-3.5m & AGN  & 0.1920 & 0.1926 & 0.1910 & 2016AJ....152...25M \\
PQSJ011745.63+363714.6 &  128.8032 &  -25.9506 & 17.51 & MDM-2.4m & AGN  & 0.1058 & 0.1062 & 0.1100 & 2015RAA....15.1438H \\
PQSJ011849.46+383618.8 &  128.7829 &  -23.9551 & 17.92 & APO-3.5m & AGN  & 0.2157 & 0.2162 & 0.2200 & 2015RAA....15.1438H \\
PQSJ012400.58+410709.9 &  129.5100 &  -21.3358 & 17.57 & APO-3.5m & AGN  & 0.3465 & 0.3490 &  &  \\
PQSJ013925.09+370454.9 &  133.4563 &  -24.8072 & 17.39 & APO-3.5m & AGN  & 0.1439 & 0.1440 &  &  \\
PQSJ021248.58+330643.3 &  141.9604 &  -26.7352 & 17.69 & APO-3.5m & AGN  & 0.1883 & 0.1887 &  &  \\
PQSJ021935.13+534019.9 &  135.8717 &   -6.9702 & 17.26 & APO-3.5m & AGN  &   & 0.0005 &  &  \\
PQSJ031304.24+461120.7 &  146.9644 &   -9.9689 & 19.75 & APO-3.5m & Star &   & 0.0000 &  &  \\
PQSJ031409.36+471111.0 &  146.5947 &   -9.0222 & 18.82 & APO-3.5m & AGN  &   & 1.2919 &  &  \\
PQSJ032839.48+451440.5 &  149.7976 &   -9.2689 & 20.14 & APO-3.5m & AGN  & 0.1038 & 0.1044 &  &  \\
PQSJ040821.86+604649.4 &  144.7800 &    6.5612 & 19.74 & APO-3.5m & Star &   & 0.0000 &  &  \\
PQSJ052113.09+802100.9 &  132.6207 &   23.0984 & 17.23 & APO-3.5m & AGN  & 0.2623 & 0.2627 &  &  \\
PQSJ053715.36+345344.8 &  174.0027 &    1.6442 & 18.29 & APO-3.5m & Star &   & 0.0000 &  &  \\
PQSJ055425.29-280544.8 &  233.4567 &  -24.0828 & 16.65 & APO-3.5m & AGN  &   & 0.0626 & 0.0630 & 2016AJ....152...25M \\
PQSJ060105.70-261108.9 &  232.0496 &  -22.0393 & 16.47 & APO-3.5m & AGN  & 0.0390 & 0.0389 & 0.0390 & 2009MNRAS.399..683J \\
PQSJ063117.46-290730.7 &  237.5212 &  -16.8695 & 18.45 & APO-3.5m & Star &   & 0.0000 &  &  \\
PQSJ064011.77-255342.1 &  235.2352 &  -13.7918 & 17.23 & APO-3.5m & AGN  &   & 0.0256 & 0.0258 & 2009MNRAS.399..683J \\
PQSJ065030.88-195929.0 &  230.7776 &   -9.1639 & 18.59 & APO-3.5m & Star &   & 0.0000 & 0.0258 & 2002LEDA.........0P \\
PQSJ072108.99-251033.6 &  238.6349 &   -5.1548 & 18.03 & APO-3.5m & Star &   & 0.0000 &  &  \\
PQSJ074522.77-135543.3 &  231.5213 &    5.2602 & 17.20 & APO-3.5m & Star &   & 0.0000 &  &  \\
PQSJ080919.34-202653.8 &  240.0980 &    6.8151 & 18.32 & APO-3.5m & AGN  &   & 0.5304 &  &  \\
PQSJ082559.42-135142.5 &  236.6484 &   13.7220 & 17.43 & APO-3.5m & AGN  & 0.3194 & 0.3199 & 0.3190 & 2016AJ....152...25M \\
PQSJ084320.25-264017.3 &  249.7434 &    9.6936 & 18.40 & APO-3.5m & AGN  &   & 0.6495 &  &  \\
PQSJ085506.79-030336.5 &  231.1207 &   25.6137 & 17.67 & APO-3.5m & AGN  & 0.2689 & 0.2694 &  &  \\
PQSJ091134.96-134801.4 &  243.3322 &   22.7939 & 17.93 & APO-3.5m & AGN  & 0.2171 & 0.2167 &  &  \\
PQSJ091945.29-063225.5 &  238.2226 &   28.7612 & 16.65 & APO-3.5m & AGN  & 0.7066 & 0.7068 & 0.7061 & 2009MNRAS.399..683J \\
PQSJ092100.39-051545.7 &  237.2411 &   29.7594 & 17.36 & APO-3.5m & AGN  &   & 0.3804 & 0.3800 & 2016AJ....152...25M \\
PQSJ092315.05-132826.5 &  244.9822 &   25.2169 & 18.18 & APO-3.5m & AGN  & 0.1877 & 0.1876 &  &  \\
PQSJ092751.79-203451.5 &  251.6378 &   21.4229 & 16.76 & APO-3.5m & Star &   & 0.0000 & 0.3474 & 2009MNRAS.399..683J \\
PQSJ093622.14-113433.9 &  245.6347 &   28.8930 & 17.78 & APO-3.5m & AGN  & 0.0912 & 0.0917 & 0.0912 & 2009MNRAS.399..683J \\
PQSJ100826.52-273544.3 &  264.0898 &   22.7249 & 17.00 & APO-3.5m & AGN  & 0.3071 & 0.3081 & 0.3080 & 2016AJ....152...25M \\
PQSJ102239.94-302930.6 &  268.6664 &   22.3055 & 17.40 & APO-3.5m & AGN  & 0.3165 & 0.3168 & 0.3165 & 2009MNRAS.399..683J \\
PQSJ110331.56-325116.7 &  278.1173 &   24.7652 & 16.91 & APO-3.5m & AGN  & 0.3554 & 0.3557 & 0.3555 & 2009MNRAS.399..683J \\
PQSJ162526.51+852941.9 &  118.7615 &   29.7461 & 17.47 & APO-3.5m & AGN  & 0.0631 & 0.0634 & 0.0629 & 2002LEDA.........0P \\
PQSJ173840.53+322409.2 &   56.9083 &   28.6152 & 16.90 & MDM-1.3m & AGN  & 0.1273 & 0.1278 & 0.1300 & 2017ApJS..233....3T \\
PQSJ174638.78+411047.3 &   67.0808 &   29.1495 & 17.27 & MDM-1.3m & AGN  & 0.3008 & 0.3010 &  &  \\
PQSJ174704.02+260336.2 &   50.7509 &   24.9571 & 18.43 & MDM-2.4m & AGN  & 0.3416 & 0.3422 & 0.3403 & 2018AJ....155..189D \\
PQSJ174707.17+383519.4 &   64.2172 &   28.4954 & 17.71 & APO-3.5m & AGN  & 0.1338 & 0.1345 &  &  \\
PQSJ175044.52+255518.0 &   50.9117 &   24.1326 & 18.47 & MDM-2.4m & AGN  &   & 0.3899 & 0.3900 & 2018AJ....155..189D \\
PQSJ180055.04+423957.1 &   69.3756 &   26.8802 & 17.14 & MDM-1.3m & AGN  & 0.1691 & 0.1694 &  &  \\
PQSJ180650.80+694928.4 &  100.1314 &   29.1668 & 17.07 & MDM-2.4m & AGN  & 0.0506 & 0.0518 & 0.0510 & 2016AJ....152...25M \\
PQSJ181234.59+395126.5 &   66.9813 &   24.0494 & 17.25 & MDM-1.3m & AGN  & 0.2420 & 0.2422 &  &  \\
PQSJ181828.96+674125.6 &   97.7368 &   28.0517 & 16.45 & MDM-1.3m & AGN  & 0.3128 & 0.3132 & 0.3140 & 2004ApJ...617..192M \\
PQSJ182408.93+464017.1 &   74.7816 &   23.8437 & 16.95 & MDM-2.4m & AGN  & 0.0670 & 0.0675 & 0.0696 & 2011MNRAS.416.2840L \\
PQSJ182446.44+650924.7 &   94.9492 &   27.2062 & 17.21 & MDM-2.4m & AGN  & 0.3044 & 0.3046 & 0.3000 & 1995A\&A...300..323J \\
PQSJ183023.19+731310.6 &  104.0389 &   27.3960 & 14.67 & APO-3.5m & AGN  &   & 0.1240 & 0.1230 & 2001AJ....121.2843B \\
PQSJ183249.71+534021.8 &   82.5863 &   24.2128 & 17.12 & MDM-2.4m & AGN  & 0.0457 & 0.0459 & 0.0390 & 2016ApJ...818..113N \\
PQSJ183503.36+324146.5 &   61.3051 &   17.4461 & 15.71 & MDM-2.4m & AGN  & 0.0587 & 0.0591 & 0.0581 & 2002LEDA.........0P \\
PQSJ183849.18+480234.4 &   76.9498 &   21.8289 & 17.21 & MDM-2.4m & Star &   & 0.0000 & 0.3000 & 2013MNRAS.430.2464M \\
PQSJ184640.81+635128.7 &   93.9346 &   24.6757 & 17.25 & APO-3.5m & AGN  & 0.4081 & 0.4087 & 0.4080 & 2016AJ....152...25M \\
PQSJ185801.14+485023.3 &   78.8222 &   19.0554 & 17.69 & APO-3.5m & AGN  & 0.0787 & 0.0790 & 0.0790 & 2012ApJ...751...52E \\
PQSJ190525.86+422739.2 &   73.0896 &   15.5315 & 17.76 & APO-3.5m & AGN  & 0.0277 & 0.0280 & 0.0275 & 2002LEDA.........0P \\
PQSJ193008.30+020449.4 &   39.2395 &   -7.6383 & 19.36 & APO-3.5m & Star &   & 0.0000 &  &  \\
PQSJ193810.75+540856.0 &   86.4448 &   15.3090 & 17.55 & APO-3.5m & AGN  & 0.2726 & 0.2731 &  &  \\
PQSJ193929.56+700748.8 &  102.0032 &   21.4616 & 17.30 & MDM-2.4m & AGN  & 0.1161 & 0.1163 & 0.1200 & 2016ApJ...818..113N \\
PQSJ194635.31+041727.5 &   43.1722 &  -10.2178 & 17.48 & APO-3.5m & Star &   & 0.0000 &  &  \\
PQSJ202610.07+112928.0 &   54.6127 &  -15.0388 & 17.04 & APO-3.5m & AGN  & 0.0861 & 0.0864 &  &  \\
PQSJ203311.82+173234.1 &   60.7865 &  -13.1253 & 17.22 & APO-3.5m & AGN  & 0.2350 & 0.2357 & 0.2340 & 2016AJ....152...25M \\
PQSJ203413.30-040529.0 &   41.3933 &  -24.6733 & 16.52 & APO-3.5m & AGN  & 0.1004 & 0.1010 &  &  \\
PQSJ204237.21+750802.1 &  108.9977 &   19.5267 & 17.23 & APO-3.5m & AGN  & 0.1042 & 0.1041 & 0.1040 & 2002LEDA.........0P \\
PQSJ205256.03+024546.2 &   50.5220 &  -25.3097 & 16.88 & MDM-2.4m & AGN  & 0.3613 & 0.3610 &  &  \\
PQSJ205820.58+340138.6 &   77.5978 &   -7.6261 & 18.62 & APO-3.5m & AGN  &   & 0.9004 &  &  \\
PQSJ210931.91+353258.5 &   80.2819 &   -8.3528 & 17.44 & APO-3.5m & AGN  & 0.2019 & 0.2023 & 0.2015 & 2013ApJ...767...14P \\
PQSJ211420.95+252422.9 &   73.2535 &  -15.8905 & 17.50 & APO-3.5m & AGN  & 0.0890 & 0.0893 &  &  \\
PQSJ211452.57+060742.9 &   57.0403 &  -28.0144 & 16.58 & MDM-2.4m & AGN  &   & 0.4608 & 0.4570 & 2017AJ....154..114O \\
PQSJ211547.08+190843.0 &   68.4845 &  -20.2226 & 16.79 & APO-3.5m & AGN  & 0.4118 & 0.4120 & 0.4100 & 2016AJ....152...25M \\
PQSJ214153.44+315127.1 &   82.4352 &  -15.7309 & 17.55 & APO-3.5m & AGN  & 0.0433 & 0.0436 & 0.0433 & 2002LEDA.........0P \\
PQSJ214335.53+174349.4 &   72.1156 &  -26.0841 & 16.92 & MDM-2.4m & AGN  & 0.2105 & 0.2112 & 0.2130 & 2016ApJ...818..113N \\
PQSJ215647.45+224250.2 &   78.4262 &  -24.6981 & 15.97 & APO-3.5m & AGN  &   & 1.3029 & 1.2900 & 2016ApJ...818..113N \\
PQSJ220314.97+314538.5 &   85.9568 &  -18.7787 & 16.24 & APO-3.5m & AGN  & 0.2949 & 0.2954 & 0.2970 & 2016ApJ...818..113N \\
PQSJ220651.79+275758.6 &   84.0498 &  -22.2169 & 17.24 & MDM-2.4m & AGN  & 0.2529 & 0.2528 &  &  \\
PQSJ221536.83+290235.4 &   86.4035 &  -22.5622 & 16.69 & MDM-2.4m & AGN  & 0.2289 & 0.2293 & 0.2290 & 1999A\&AS..139..575W \\
PQSJ223954.44+272431.9 &   90.1932 &  -26.9743 & 17.38 & APO-3.5m & AGN  & 0.0650 & 0.0656 & 0.0654 & 2020ApJS..250....8L \\
PQSJ224436.88+475301.3 &  101.9768 &   -9.7610 & 18.83 & APO-3.5m & AGN  &   & 0.8993 &  &  \\
PQSJ225147.70+341930.6 &   96.4956 &  -22.3319 & 17.42 & MDM-2.4m & AGN  & 0.1318 & 0.1322 & 0.1320 & 1999A\&AS..139..575W \\
PQSJ225603.39+273209.7 &   93.7370 &  -28.7147 & 17.07 & APO-3.5m & AGN  & 0.3632 & 0.3641 & 0.3635 & 2020ApJS..250....8L \\
PQSJ234229.32+530025.5 &  112.5463 &   -8.4685 & 19.23 & APO-3.5m & AGN  & 0.1520 & 0.1524 &  &  \\
\enddata
\tablecomments{PQS candidates for which we obtained APO or MDM spectra. In addition to their PQS name, Galactic longtitude and latitude, and NUV magnitude, we provide the name of the telescope at which the data were taken, the spectral classification (either Star or AGN). We provide two redshift estimates for the confirmed AGN when possible: (1) z$_{\rm narrow}$ is estimated based on any narrow emission lines present in the spectrum, typically [OIII]  $\lambda \lambda$ 4960, 5008. (2) z$_{\rm MARZ}$ is the redshift based on QSO template fitting by the Manual and Automated Redshifting Software \citep{Hinton16} or MARZ. When available, we also list previously published spectroscopic redshifts for the candidates, along with their sources.}
\end{deluxetable*}

\bibliographystyle{aasjournal}
\bibliography{planeqsorefs}

\end{document}